\documentclass[10pt,letterpaper]{article}
\usepackage[top=0.85in,left=2.75in,footskip=0.75in]{geometry}

\usepackage{amsmath,amssymb}

\usepackage{changepage}

\usepackage[utf8x]{inputenc}

\usepackage{textcomp,marvosym}

\usepackage{cite}

\usepackage{nameref,hyperref}

\usepackage[right]{lineno}

\usepackage{microtype}
\DisableLigatures[f]{encoding = *, family = * }

\usepackage[table]{xcolor}

\usepackage{array}
\usepackage{graphicx}%
\usepackage{multirow}%
\usepackage{amsmath,amssymb,amsfonts}%
\usepackage{amsthm}%
\usepackage{mathrsfs}%
\usepackage[title]{appendix}%
\usepackage{xcolor}%
\usepackage{textcomp}%
\usepackage{manyfoot}%
\usepackage{booktabs}%
\usepackage{subfig}
\usepackage{algorithm}%
\usepackage{algorithmicx}%
\usepackage{algpseudocode}%
\usepackage{listings}%
\usepackage{graphbox}
\usepackage{enumitem}
\usepackage{wrapfig}
\usepackage{lineno}
\usepackage{soul}
\usepackage{tcolorbox}
\usepackage{enumerate}
\usepackage{colortbl}

\newcolumntype{+}{!{\vrule width 2pt}}

\newlength\savedwidth

\raggedright
\usepackage[aboveskip=1pt,labelfont=bf,labelsep=period,justification=raggedright,singlelinecheck=off]{caption}

\makeatletter
\renewcommand{\@biblabel}[1]{\quad#1.}
\makeatother

\usepackage{lastpage,fancyhdr,graphicx}
\usepackage{epstopdf}
\fancyheadoffset[L]{2.25in}
\fancyfootoffset[L]{2.25in}
\newcommand{\added}[1]{\textcolor{black}{#1}}

\newcommand{\raceethnicity}[0]{\textsc{Race and Ethnicity }}
\newcommand{\urbanizationlevel}[0]{\textsc{Urbanization Level} }

\begin{document}
\vspace*{0.2in}

\begin{flushleft}
{\Large
\textbf\newline{Auditing the Fairness of the US COVID-19 Forecast Hub's Case Prediction Models} % Please use "sentence case" for title and headings (capitalize only the first word in a title (or heading), the first word in a subtitle (or subheading), and any proper nouns).
}
\newline
\\
Saad Mohammad Abrar\textsuperscript{1},
Naman Awasthi\textsuperscript{1},
Daniel Smolyak\textsuperscript{1},
Nekabari Sigalo\textsuperscript{3},
Vanessa Frias Martinez\textsuperscript{1,2}
\\
\bigskip
\textbf{1} Department of Computer Science, University of Maryland, College Park, MD, USA\\
\textbf{2} College of Information and UMIACS, University of Maryland, College Park, MD, USA\\
\textbf{3} Abt Global\\
%\\
%\textbf{2} Affiliation Dept/Program/Center, Institution Name, City, State, Country
%\\
%\textbf{3} Affiliation Dept/Program/Center, Institution Name, City, State, Country
%\\
\bigskip

% Insert additional author notes using the symbols described below. Insert symbol callouts after author names as necessary.
% 
% Remove or comment out the author notes below if they aren't used.
%
% Primary Equal Contribution Note
%\Yinyang These authors contributed equally to this work.

% Additional Equal Contribution Note
% Also use this double-dagger symbol for special authorship notes, such as senior authorship.
%\ddag These authors also contributed equally to this work.

%\textcurrency Current Address: Dept/Program/Center, Institution Name, City, State, Country % change symbol to "\textcurrency a" if more than one current address note
% \textcurrency b Insert second current address 
% \textcurrency c Insert third current address


% Group/Consortium Author Note
%\textpilcrow Membership list can be found in the Acknowledgments section.

% Use the asterisk to denote corresponding authorship and provide email address in note below.
%* sabrar@umd.edu

\end{flushleft}
\section*{Abstract}
The \added{US} COVID-19 Forecast Hub, a repository of COVID-19 forecasts from over 50 independent research groups, is used by the Centers for Disease Control and Prevention (CDC) for their
official COVID-19 communications. As such, the Forecast Hub is a critical centralized resource to promote transparent decision making. \added{While the Forecast Hub has provided valuable predictions focused on accuracy, there is an opportunity to evaluate model performance across social determinants such as race and urbanization level} that have been known to play a role in the COVID-19 pandemic. In this paper, we carry out a comprehensive fairness analysis of the Forecast Hub model predictions and we show statistically significant diverse predictive performance across social determinants, with minority racial and ethnic groups as well as less urbanized areas often associated with higher prediction errors. We hope this work will encourage COVID-19 modelers and the CDC to report fairness metrics together with accuracy, and to reflect on the potential harms of the models on specific social groups and contexts.

% Please keep the Author Summary between 150 and 200 words
% Use first person. PLOS ONE authors please skip this step. 
% Author Summary not valid for PLOS ONE submissions.   
%\section*{Author summary}
%Lorem ipsum dolor sit amet, consectetur adipiscing elit. Curabitur eget porta erat. Morbi consectetur est vel gravida pretium. Suspendisse ut dui eu ante cursus gravida non sed sem. Nullam sapien tellus, commodo id velit id, eleifend volutpat quam. Phasellus mauris velit, dapibus finibus elementum vel, pulvinar non tellus. Nunc pellentesque pretium diam, quis maximus dolor faucibus id. Nunc convallis sodales ante, ut ullamcorper est egestas vitae. Nam sit amet enim ultrices, ultrices elit pulvinar, volutpat risus.

% Use "Eq" instead of "Equation" for equation citations.
\section*{Introduction}\label{sec1}
\subsection*{The \added{US} COVID-19 Forecast Hub: Development and Context}
The \added{US} COVID-19 Forecast Hub was founded in 2020 and serves as a \textit{``central repository of COVID-19 forecasts from over 50 independent research groups"} \cite{forecasthub}. 
Participant research groups submit county, state and national US COVID-19 %predictions 
%for future days or weeks 
forecasts with a standardized format; and the Forecast Hub provides an interactive visualization tool to help decision makers and the general public analyze weekly predictions for COVID-19 hospitalizations, cases and deaths. The standardized predictions collected from all research groups, as well as the predictions for an ensemble model that brings all individual predictions together, are also shared with the Centers for Disease Control and Prevention (CDC) who uses these results for their official COVID-19 communications~\cite{cdc}. 

Over the past four years, numerous research groups from both academia and industry have focused on developing models to forecast COVID-19 cases, hospitalizations and deaths in the United States. The COVID-19 Forecast Hub \cite{Cramer2022-hub-dataset} has been instrumental in collating these efforts.
%showcasing predictions from over 50 teams at both state and county levels, including incremental case forecasts. These models vary in approach, 
These models vary in approach, 
ranging from deep learning methods \cite{zhang2021seq2seq,arik2020interpretable, lucas2023spatiotemporal, le2020neural} to compartmental models \cite{pei2020initial, zou2020epidemic}, statistical models \cite{chiang2022hawkes,galasso2022random}, or combinations of these via ensemble models 
%. Notably, ensemble models, both hub-wide and multi-model, have also been prevalent 
\cite{adiga2023phase, Cramer2022-hub-dataset}. COVID-19 forecast models are usually trained with historical data (e.g., past cases or hospitalizations) together with other contextual information such as human mobility data. 
Human mobility data has been used in the past to model and characterize human behaviors in the built environment 
~\cite{vieira2010querying,hernandez2017estimating,frias2013cell,rubio2010human,wuspatial}, to support decision making for socio-economic development
~\cite{frias2010socio,fu2018identifying,frias2012mobilizing, hong2016topic,frias2012computing}, for public safety~\cite{wu2022enhancing,wu2023auditing}, as well as during epidemics and 
disasters~\cite{wesolowski2012quantifying,bengtsson2015using,hong2017understanding,isaacman2018modeling,ghurye2016framework,hong2020modeling}. During the COVID-19 pandemic, human mobility has also played a central role in driving decision making, and
%for example, with social distancing policies significantly reducing the spread of the virus ~\cite{badr2020association}.
more than $50\%$ of the Forecast Hub models have incorporated mobility data into their prediction models, acknowledging the impact of human movement on virus propagation \cite{arik2020interpretable, lucas2023spatiotemporal, le2020neural, erfani2023fairness, badr2021limitations, abrar2023analysis}. 

\subsection*{Fairness in COVID-19 Prediction Models: A Critical Gap}

The \added{US} COVID-19 Forecast Hub has been, and continues to be, a critical centralized resource to promote transparent decision making. 
%where different research groups share COVID-19 hospitalizations, cases and deaths predictions in a standardized way. By allowing to compare and combine predictions across models, the Forecast Hub promotes and facilitates the use of COVID-19 predictions for decision making and the general public in a transparent way. That effort is laudable. 
\added{While the Forecast Hub has made significant contributions through its accuracy-focused predictions at different spatial granularities (\textit{e.g.,} county or state), there is an opportunity to expand its evaluation framework to examine how prediction performance varies across social determinants like race, ethnicity and urbanization levels that have been shown to play an important role in COVID-19, including race, ethnicity and rurality~\cite{gross2020racial,souch2021commentary}.}

%Nevertheless, by focusing exclusively on prediction accuracy at different spatial granularities (\textit{e.g.,} county or state), 
%the Forecast Hub fails to evaluate whether the proposed models are fair \textit{i.e.,} share similar prediction performance across social determinants that have been known to play a role in COVID-19 including race, ethnicity and rurality~\cite{gross2020racial,souch2021commentary}.

The pandemic has highlighted existing disparities in healthcare, with significant differences in COVID-19 infection rates, hospital admissions, and deaths among different racial and ethnic groups as well as across the urban-rural spectrum \cite{tsai2022algorithmic, souch2021commentary}. These disparities risk being perpetuated in model predictions if not adequately addressed. 
Diverse prediction performance across social determinants - for example, higher prediction errors for a given minority race or ethnicity - could negatively impact 
resource allocation and intervention decisions e.g., hospital beds or stay-at-home orders,
%how resources are allocated for social groups that might need the most support, 
given that the CDC appears to be using the Forecast Hub predictions for official communications that subsequently inform policy decisions~\cite{cdc}. \added{Given the urgent need for rapid pandemic response modeling, initial Forecast Hub efforts necessarily focused on developing accurate predictions. As these models continue to inform CDC communications and policy decisions, incorporating fairness analyses could further enhance their utility for equitable resource allocation and intervention planning across diverse communities ~\cite{rajkomar2018ensuring}}.

There are many reasons why the COVID-19 prediction performance can be different across social determinants such as race, ethnicity or urbanization levels. The Forecast Hub's COVID-19 prediction models are trained on datasets containing COVID-19 statistics for hospitalizations, cases or deaths. \added{Given the unprecedented scale and urgency of the pandemic, data collection faced several challenges}~\cite{douglas2021variation,gross2020racial}. For example, a lack of consistency in reporting race and ethnicity across jurisdictions, has generated a lot of missing racial data. That data is often excluded due to its incompleteness, potentially affecting the actual total hospitalizations, cases or deaths for minority race and ethnicity groups which might be less reported. 
%BUT, cases that do not have a race are NOT removed from total counts hence it does not affect general predictions...and race being wrong also does not affect total cases...we have to argue cases are under-reported. 
In addition, there are occasions where the race is reported by the medical staff instead of being self-reported, which is the most accurate source and prevents errors \added{\cite{boehmer2002self}}. For example, the CDC reports that the latest research on race and Hispanic origin misclassification on COVID-19 death certificates shows that deaths are underreported by 33\% for non-Hispanic American Indian or Alaska Natives, by 3\% for non-Hispanic Asian or Pacific Islanders, and by 3\% for Hispanic decedents~\cite{cdcunderreporting}.
%under-reporting https://www.cdc.gov/nchs/nvss/vsrr/covid19/health_disparities.htm#CountyRaceHispanicOrigin
Testing availability and access varied across communities, with some minority groups experiencing more limited access, such as Latino communities ~\cite{ama}, thus affecting the accuracy of the overall COVID-19 statistics, with under-reporting bias perpetuating the invisibility of racial and ethnic minorities in general COVID-19 statistics. A similar effect has been observed in rural counties and states, with rural areas associated to lower testing rates, thus disproportionately detecting fewer cases of COVID-19 in these regions~\cite{souch2021commentary}.
%suggesting that the least healthy, rural states have the poorest rates of COVID‐19 testing. 12 We offer a complementary commentary on the surveillance of COVID‐19 and its risk factors in rural populations. Our analyses suggest that rural states—ranked higher in specific risk factors like hypertension, 1 obesity, 1 diabetes, 1 lung cancer, 1 and e‐cigarette use 2 —are performing tests at lower rates. Moreover, we find that despite these vulnerabilities, rural states are detecting disproportionately fewer cases of COVID‐19.
                    
To exacerbate this situation even more, COVID-19 prediction performance across social determinants can also be affected by additional datasets used in the training of some of the COVID-19 prediction models. Specifically, around
50\% of the Forecast Hub's models use human mobility data from Safegraph~\cite{safegraph}, Apple~\cite{apple} or Google~\cite{google} among others, to complement COVID-19 predictions (see Fig \ref{fig:2a}). Human mobility data can characterize origin-destination trips, visits to specific points of interest (POI), or the volumes of different types of trips (\textit{e.g.,} car vs. public transit). Research has shown that mobility data can improve the prediction accuracy of COVID-19 cases, deaths and hospitalizations~\cite{ilin2021public, abrar2023analysis, garcia2021improving}. Nevertheless, researchers have also identified that mobility data suffers from sampling bias across race and age groups~\cite{erfani2023fairness} with, for example, elder and Black communities being less represented~\cite{coston2021leveraging}. Similarly to the COVID-19 case under-reporting bias, mobility data sampling bias could also affect the fairness of COVID-19 predictions across social groups.

In this paper, we propose - to the best of our knowledge - the first thorough fairness analysis of the COVID-19 prediction models in the Forecast Hub. Specifically, we focus on COVID-19 case prediction models at the county level, since these are closer to local realities and allow for more actionable decision making than state-level predictions. 
We use error parity as a measure of group fairness ~\cite{gursoy2022error} \textit{i.e.,} lack of fairness in our context is associated with significantly different error distributions across two 
%to reveal whether the distribution of prediction errors differs significantly across groups defined by two 
social determinants: race or ethnicity and urbanization level.
%Prior work has proposed several regression fairness metrics to assess group fairness in regression-type models like the ones used for COVID-19 predictions
%We propose to measure the group fairness of COVID-19 case prediction 
%models by assessing the differences in error rates at the county level across two protected attributes: race and ethnicity, and rurality~\
%In the context of Covid-19 prediction, falls into the notion of group fairness \cite{mehrabi2021survey}, and given the nature of the quantile predictions, is a fair regression task, rather than a classification fairness task. Particularly, statistical parity is very relevant, where predictions are expected to be statistically independent of the protected attribute. There has been a glut of work focusing entirely on defining fairness metrics in these settings 
%~\cite{agarwal2019fair,calders2013controlling, chzhen2020fair};
%and researchers have proposed multiple hypothesis testing frameworks to identify the presence of statistically significant differences in fairness metrics across various social determinants  
%Nevertheless, all these works fail to essentially capture objective thresholds which are needed to differentiate reasonable differences from evidence of bias. Given that, multiple hypothesis testing frameworks 
%~\cite{si2021testing, gursoy2022error, taskesen2021statistical, blanchet2021statistical,tramer2017fairtest}.
Accurately revealing differences across racial and ethnic groups would require access to county-level COVID-19 case data 
%To compute the equalized odds of a model across race, ethnicity and rurality, would require having access to daily, county COVID-19 case data 
stratified by race or ethnicity,
%and urbanicity levels,
%and rurality levels, 
which would allow us to compare predicted versus actual case county statistics for each racial and ethnic group.
%and urbanicity level. 
%across the two protected attributes. 
\added{Due to the complexity and scope of pandemic data collection efforts, many counties in the US faced significant challenges in collecting comprehensive demographic data~\cite{kader2021participatory}}. Hence, to be able to carry out a fairness analysis of the Forecast Hub's COVID-19 prediction models, we propose \added{a regression analysis to evaluate the associations between prediction errors in a given county and the race and ethnicity distributions for that county, while controlling for underlying health conditions and age groups. A similar regression analysis is proposed for the urbanization levels.}

%several approaches to associate prediction errors in a given county with race or ethnicity 
%and rurality 
%labels using information from the American Community Survey (ACS)~\cite{acs}; and evaluate whether these race association approaches might impact the fairness analysis. 

Additionally, to support researchers in the Forecast Hub, we also investigate how group fairness metrics for race, ethnicity and urbanicity levels change across model characteristics such as model type (\textit{e.g.,} deep learning versus statistical), training data (\textit{e.g, with or without mobility data}), lookaheads (\textit{e.g.,} predicting cases for next week versus in four weeks) or pandemic phases. 
Finally, we also describe a dashboard that we have designed to allow decision makers and researchers explore \textit{fairness nutritional cards} for each Forecast Hub model~\cite{stoyanovich2019nutritional}.
%In the COVID-19 county case prediction setting, we posit that we expect 
%the COVID-19 models in the Forecast Hub to have similar errors independently of race, ethnicity and rurality.
%In this paper, we carry out fairness analyses for race, ethnicity and rurality across types of models, training datasets %(w/out mobility), prediction lookaheads (1, 2, 3 or 4 weeks aheads) and COVID-19 phases; 
%and we describe a dashboard that will allow researchers and decision makers to explore 
 To sum up, the main contributions 
 %and findings 
 of this paper are:

\begin{itemize}%[leftmargin=*]

\item \added{We present a thorough fairness analysis of the CDC Forecast Hub's COVID-19 county case prediction models across race, ethnicity and urbanization levels. Our research shows statistically significant differences in predictive errors with some minority racial and ethnic groups as well as
%Black, Hispanic, Non-white groups and 
less urbanized areas associated with significantly higher errors than the majority White race, while controlling for underlying health conditions, age groups and state.} 

%some minority racial and ethnic counties
%Asian, Black, Hispanic and Non-white counties 
%as well as less urbanized counties often associated with statistically significant higher prediction errors. 

\item \added{We carry out interaction analyses identifying differences in performance 
across racial/ethnic groups and urbanicity levels with respect to COVID-19 prediction type models, COVID-19 datasets, prediction lookaheads and COVID-19 phases, while controlling for underlying health conditions and age groups. Our results show significant prediction performance differences for certain minority groups and less urbanized areas, when compartmental or statistical models are used. On the other hand, short-term forecasting and certain pandemic phases with higher case volumes are also associated with higher prediction performance differences for certain minority racial and ethnic groups as well as for less urbanized areas. }

%We propose several approaches to associate county model prediction errors to race or ethnicity, and evaluate whether different associations produce similar or different fairness results. Our analysis depicts similar findings across approaches, with higher prediction errors often associated with minority groups independently of the race association approach used. 

\item We present a dashboard where researchers and decision makers at the CDC and beyond will be able to explore 
%explain going from PBL to AER
fairness nutritional cards per individual model across race, ethnicity and urbanization level, and how fairness might vary across model and data characteristics. 
%across lookaheads and phases. 

\end{itemize}
\section*{Materials and Methods}
%\subsection*{Data Sources and Study Period}
\subsection*{Data Sources and Variables}

\subsubsection*{COVID-19 Forecast Data}
%The COVID-19 Forecast Hub \cite{Cramer2022-hub-dataset}, spearheaded by the University of Massachusetts, and used by the CDC for decision making, serves as a comprehensive repository for COVID-19 forecasts in the United States, consolidating state-level and county-level predictions. These forecasts, contributed by a diverse array of independent modeling groups, are unified within the hub for public dissemination and comparative analysis. 
For the purpose of our study, we focus exclusively on the weekly, county-level COVID-19 case predictions publicly available from the COVID-19 Forecast Hub across all US counties~\cite{forecasthub}. 
We focus on county-level forecasts because these are closer to local realities and allow for more actionable decision-making than state-level predictions. 
%\textcolor{red}{Saad, can you please confirm the next sentence is correct? or are there hospitalizations and deaths at the county level?}  %\saad{ (added to the text) Refer to Table 1 \cite{Cramer2022-hub-dataset}}
On the other hand, since hospitalizations and deaths are only available at the state level \cite{Cramer2022-hub-dataset}, we focus on COVID-19 case predictions. 
%\saad{We seek to focus only on county-level forecasts because it can provide highly localized and granular analysis, which perfectly suits our fairness assessment. State or higher level aggregations would mask the important distinctions, potentially hiding disparities that exist at more local levels.} 
The weekly incidence predictions in the Forecast Hub are uploaded by participating teams and defined as the newly anticipated 
COVID-19 cases per county within the following \textit{epidemiological week}, extending from Sunday to Saturday.  
We use the weekly forecasts during the period from July 2020 to October 2022.
%\textcolor{red}{Saad, are these dates correct?}
%We use forecasts for N ranging from 1 to 4 weeks (a.k.a lookaheads), allowing us to evaluate differences in the predictive accuracy of the hub's models for short and medium-term horizons. 

The hub's data repository offers both point forecasts and quantile-based probabilistic forecasts. Our study employs the latter, leveraging the seven provided quantiles ([0.025, 0.100, 0.250, 0.500, 0.750, 0.900, 0.975]) to gain insights into the uncertainty ranges and confidence intervals posited by the forecasting models. 
From the entire cohort of models and teams contributing to the Forecast Hub, we selected $36$ teams that met our inclusion criteria: they provided comprehensive quantile forecasts throughout our period of analysis and they submitted predictions at the county-level. 
%\textcolor{blue}{Figure reference to appendix missing in next sentence.}
A Gantt chart depicting the specific quantile forecasts used to evaluate each model is shown in Fig 1 in \nameref{s1appendix}. % \nameref{fig:gantt}). 

\begin{figure}[!ht]
\centering
\subfloat[\centering Distribution of Racial Demographics Across US Counties\label{fig:race_perc_dist}]{\includegraphics[width=0.45\textwidth]{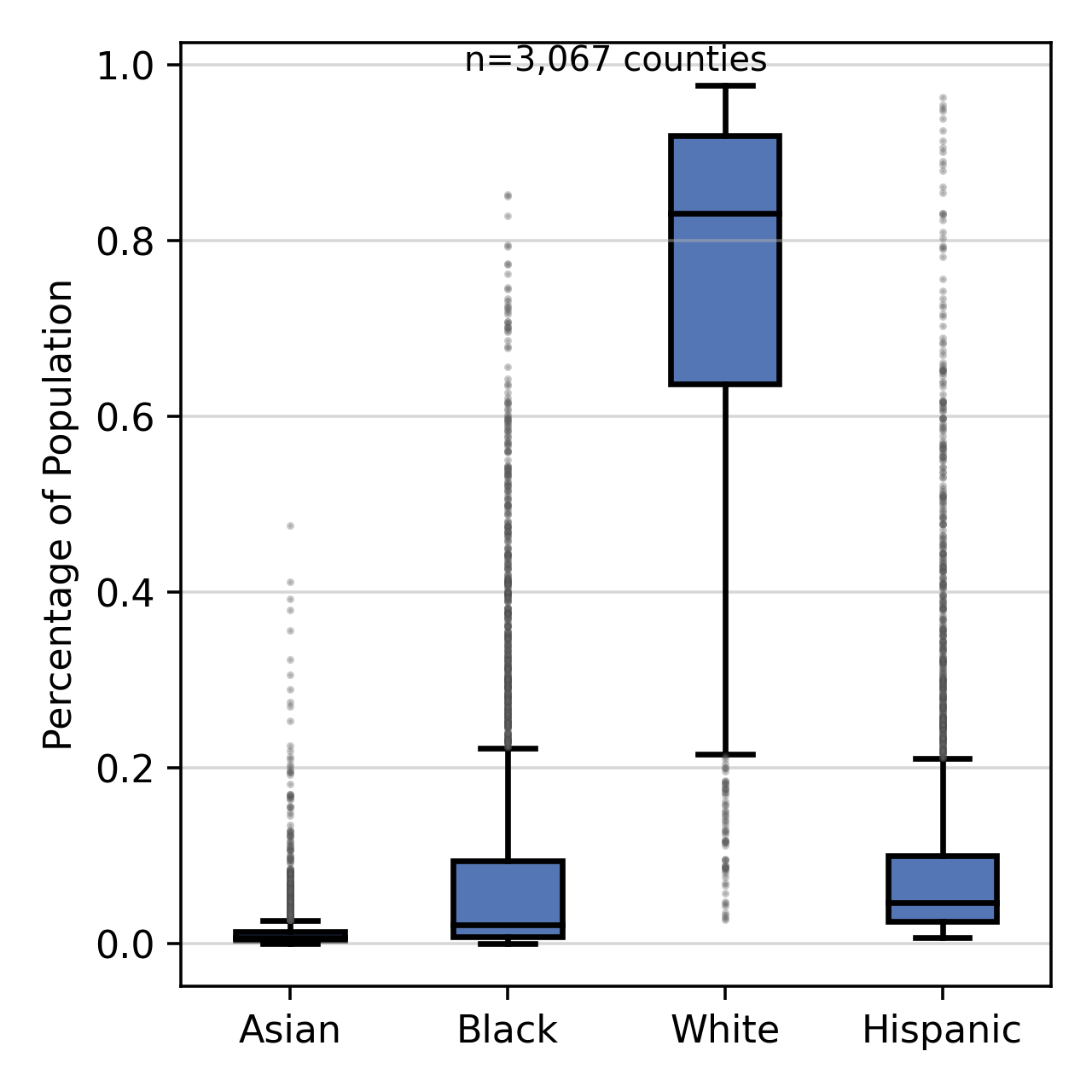}}
\centering
\subfloat[\centering Distribution of Counties by Urbanicity Level\label{fig:urbanicity_dist}]{\includegraphics[width=0.45\textwidth]{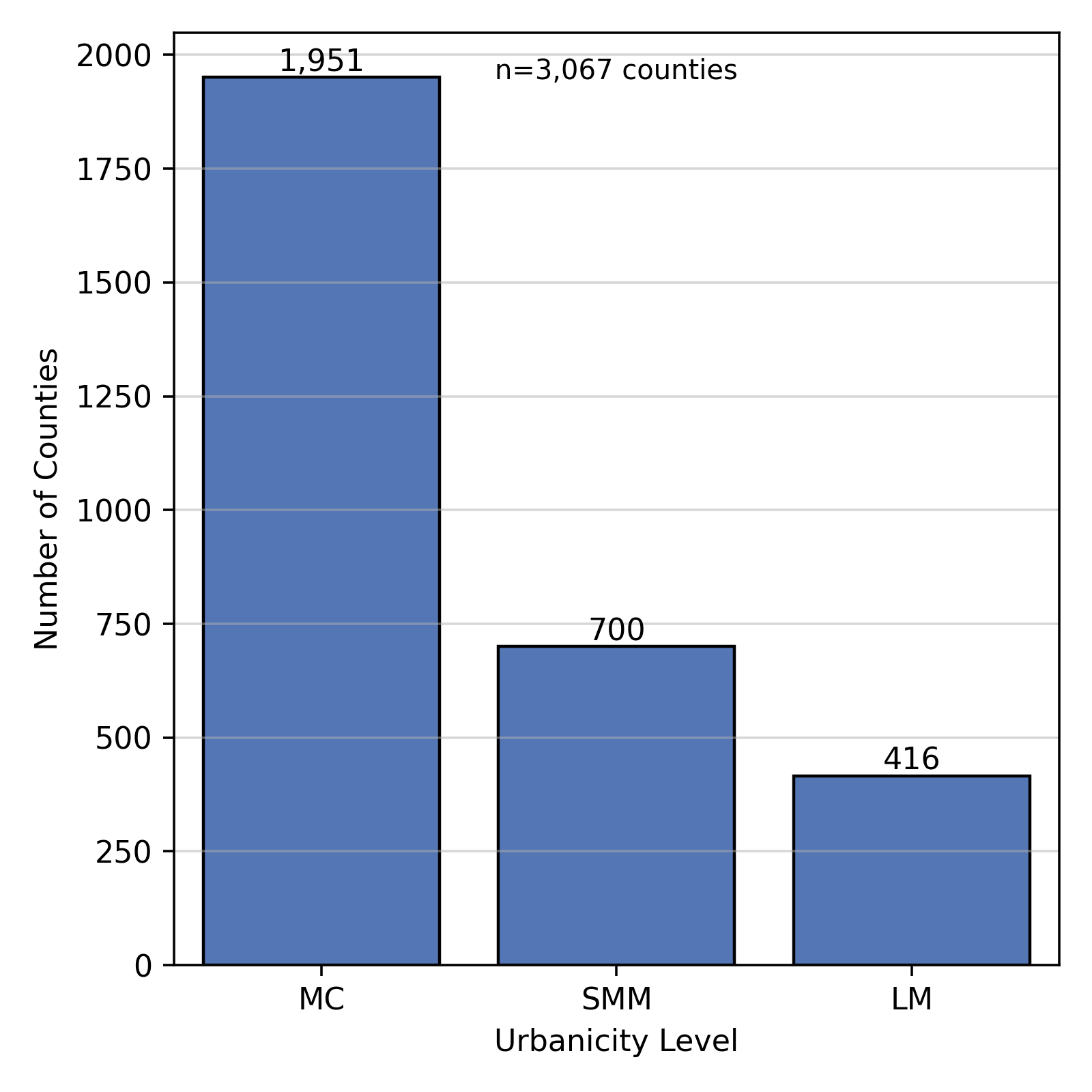}}
\caption{\centering Distribution of the sensitive attributes (Race/Ethnicity and Urbanization level) across the 3,067 counties in US considered for this study.}
\label{fig:sensitive_attributes}
\end{figure}

%\vanessa{Saad, please add color to the plots in Fig 1 (similar to Fig 2). Otherwise it looks weird. Thx!}

\subsubsection*{Sensitive Attributes}
\added{To empirically evaluate fairness in COVID-19 forecasts, we measure error parity across two sensitive demographic attributes: (1) \textsc{race and ethnicity} population composition, and (2) county-level \textsc{urbanization level} classification.  For this analysis, we use two primary data sources: the American Community Survey (ACS) \cite{acs} for demographic information and the CDC urbanization classification for county-level urban-rural designations \cite{cdcurban}}.  
%These groups represent a relatively small proportion of the U.S. population in most counties, with highly skewed geographical distributions. For instance, AIAN populations are concentrated in specific regions, particularly in tribal areas, making county-level analyses less reliable for general inference. 
\added{In our analysis, we focus on Asian, Black, Hispanic, and White racial/ethnic groups while using White as the reference category (see Fig \ref{fig:race_perc_dist} for the racial and ethnic distribution across the 3,067 counties considered for this study). Other racial categories, including American Indian/Alaska Native (AIAN), Native Hawaiian/Pacific Islander (NHPI), and multi-racial groups, were excluded from our primary analysis due to limited variable distribution. As Table \ref{tab:county_demographics} shows, 97.69\%
of counties have NHPI populations under 1\%, and 79.03\% have AIAN populations under 1\%, providing insufficient
variation for meaningful regression analysis.
In contrast, between 1.21\%-66.22\% of the counties have less than 1\% of Asian, Black, or Hispanic population, providing adequate statistical power for detecting potential effects in our regression analysis. Furthermore, Asian populations represent a key demographic group in metropolitan
and suburban areas where COVID-19 impacts were particularly pronounced, making their inclusion crucial for understanding prediction accuracy in these important contexts. Further details about this selection, and exclusion of some counties can be found in the \nameref{s1appendix} in Section 1.1. }

%%Demographic Distribution
\begin{table}[h]
\caption{County-Level Demographic Distribution Statistics}
\label{tab:county_demographics}
\resizebox{\linewidth}{!}{
\begin{tabular}{lcccc}
\hline
\textbf{Racial/Ethnic Group} & \textbf{Median (\%)} & \textbf{Mean (\%)} & \textbf{Counties $>1$\%} & \textbf{\% Counties $>1$\%}  \\
\hline
White & 83.09 & 75.64 & 3,067/3,067 & 100.00  \\
Hispanic & 4.64 & 9.88 & 3,030/3,067 & 98.79  \\
Black & 2.14 & 8.88 & 2,019/3,067 & 65.83  \\
Asian & 0.70 & 1.53 & 1,036/3,067 & 33.78  \\
American Indian/Alaska Native & 0.40 & 2.00 & 643/3,067 & 20.97  \\
Native Hawaiian/Pacific Islander & 0.12 & 0.24 & 71/3,067 & 2.31  \\
\hline
\multicolumn{5}{p{0.95\linewidth}}{\scriptsize \textbf{Notes:} Statistics are computed across all 3,067 U.S. counties included in our analysis. "Counties $>1\%$" shows the number of counties where the group comprises more than $1\%$ of the population. } \\
\end{tabular}}

\end{table}

\added{On the other hand, the CDC urban-rural classification scheme classifies counties into six urbanization levels\cite{cdcurban}, from highly urban (1) to rural (6). For this paper, we group them into three labels: 
%USDA rural-urban continuum codes that assign a given county to a code representing its degree of rurality~\cite{usda}. For this analysis, we consider the following codes: 
Large Metropolitan areas (\textbf{LM}, which correspond to codes 1 and 2),  Small and Medium Metropolitan (\textbf{SMM}, codes 3 and 4) and Micropolitan  and Non-core areas (\textbf{MC}, codes 5 and 6). 
This grouping ensures sufficient sample sizes and variations within each category for robust statistical analysis, given the uneven distribution of US counties across urbanization levels (see Fig \ref{fig:urbanicity_dist}). At the same time the three-level classification provides a clearer narrative about urban-rural disparities while maintaining meaningful distinctions in population density and healthcare infrastructure. }

%\subsubsection*{Ground Truth and Demographic Data}
%The empirical evaluation of the group fairness of these forecasts - modeled as error parity across social \added{or sensitive groups} - is conducted using the ground truth case data sourced from the JHU CSSE COVID-19 Data \cite{dong2020interactive}; race and ethnicity data from the ACS~\cite{acs} and urbanization levels from the CDC~\cite{cdcurban}.

%\textcolor{red}{Saad, add Gantt in appendix and update figure reference here}

\subsubsection*{Model-Data Characteristics}
Given our interest in understanding how group fairness metrics for race, ethnicity and urbanization might change across model and data characteristics, we break down the prediction performance and fairness analyses across four aspects: \begin{itemize}
    \item \textbf{Model Type}: Based on information reported in the papers associated to each of the $36$ predictive models, we have manually classified them into five categories, namely: Statistical, Compartmental, Deep Learning, Baseline, and Ensemble \added{(see Fig \ref{fig:2a} for model statistics).  This classification aims to discern the potential influence of model typologies on forecast performance and to identify any systematic biases inherent to specific modeling approaches (Section 1.2 and Table 1 in \nameref{s1appendix} provide more details about the taxonomy).}
    %\todo{Add justifications for the model type classification.}
    %\textcolor{red}{saad: we need to include the list of models and their labels in the appendix for transparency.}

\begin{figure}[!ht]
\centering
\subfloat[\centering Model types and mobility used counts\label{fig:2a}]{\includegraphics[width=0.5\textwidth]{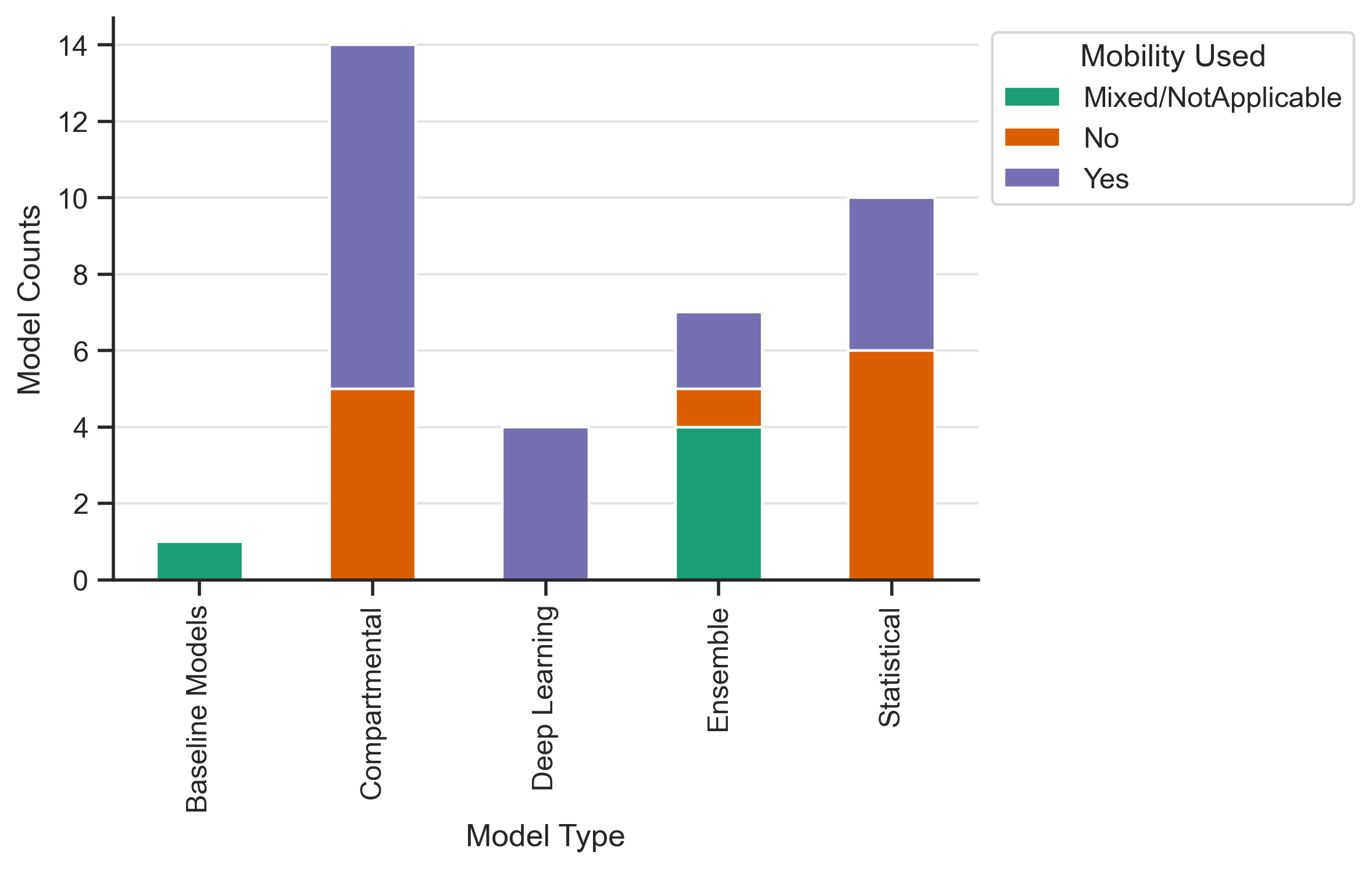}}
\centering
\subfloat[\centering Phase demarcation where P0 represents Phase 0\label{fig:2b}]{\includegraphics[width=0.5\textwidth]{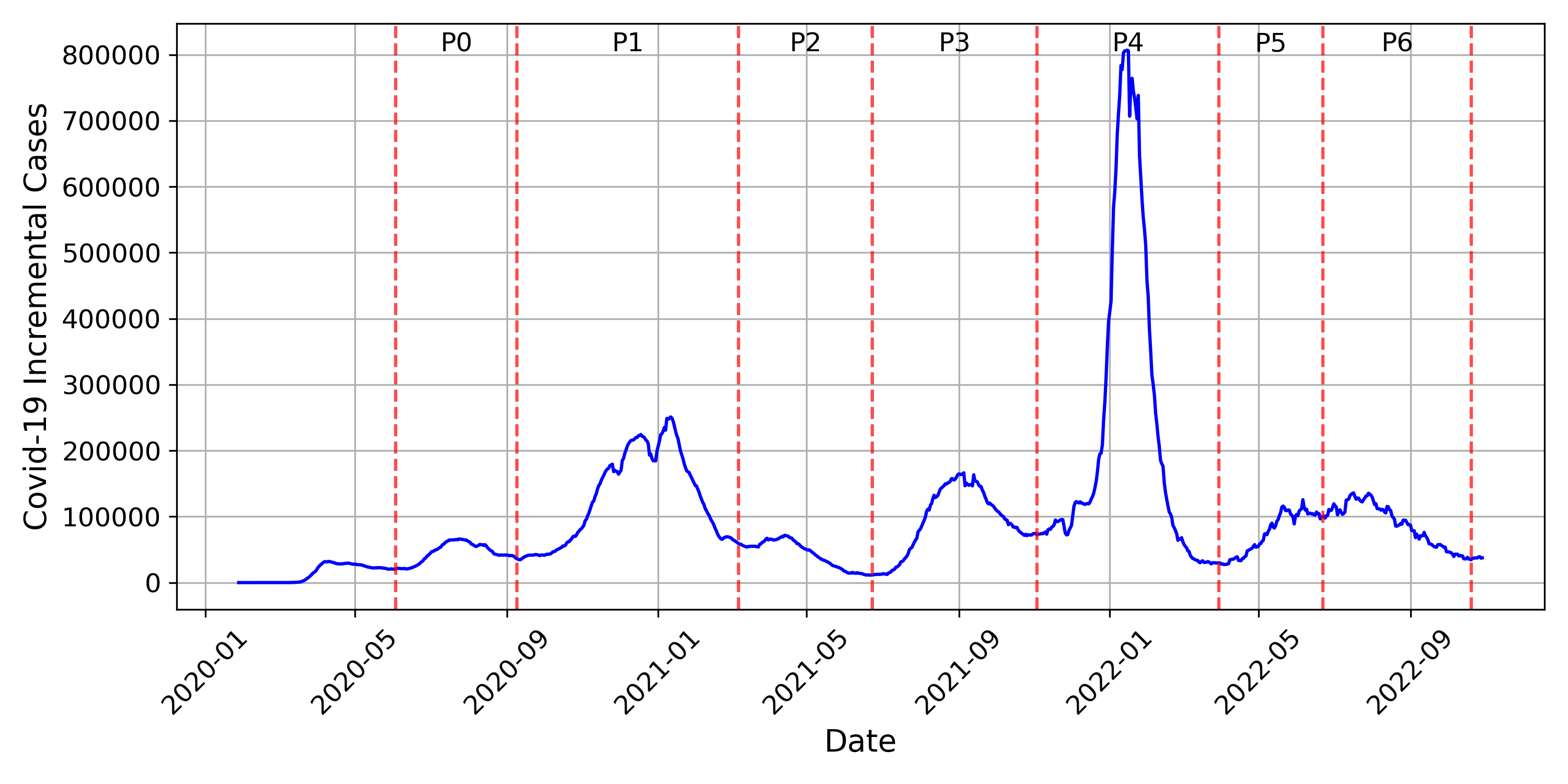}}
\caption{Prediction fairness will be evaluated across types of models, training datasets (mobility), lookaheads and phases. }
\label{fig0}
\end{figure}
    
    \item \textbf{Mobility Used}: Based on information reported in the papers for each predictive model analyzed, we distinguish between models that integrate mobility data and those that do not. The main objective of this feature is to allow us to explore the effect of mobility data on forecast accuracy and the possible introduction of biases resulting from this additional data dimension. Fig \ref{fig:2a} displays a categorization of the forecasting models, differentiated by both the type of model and the incorporation of mobility data, as quantified by their respective counts. \added{\textit{Mixed models} are COVIDHub Ensemble models (like COVIDhub CDC-ensemble) that combine models that use (or not) mobility data in their training. }

%\vanessa{Saad, I have renamed everything from N/A to Mixed, please do a second pass and make sure I changed this everywhere.} \saad{@Vanessa, Cannot be renamed to Mixed, because for example, COVIDhub-baseline ``For predictions of incidence, the median prediction at all future horizons is the most recent observed incidence. " Here: https://zoltardata.com/model/302 }

   % \item \textbf{Lookaheads:} We use forecasts for N ranging from 1 to 4 weeks (a.k.a lookaheads),
     \item \textbf{Lookahead}: We use forecasts ranging from 1 week to 4 weeks (a.k.a lookaheads), allowing us to evaluate differences in the predictive accuracy of the hub's models across race, ethnicity and the rural-urban spectrum for short and medium-term horizons: at 7, 14, 21 and 28 days.
     
     \item \textbf{Phase}: COVID-19 case numbers differ a lot across pandemic stages. To understand whether phases play a role in model fairness across race, ethnicity and urbanization level, we divide the period under study into seven phases, identified based on the %alternative 
     presence of valleys and peaks in the volume of COVID-19 cases (see Fig \ref{fig:2b}). \added{A more detailed explanation of the phase identification process can be found in Section 1.3 in the \nameref{s1appendix}. }
     %\todo{a more elaborate explanation of phases}
     %(see Figure  \ref{fig:2b}). 
     %(see Table \ref{tab:phases}). 
     %These phases are demarcated by significant troughs and peaks in case numbers, allowing us to contextualize the forecasts within specific periods of the pandemic's trajectory. 
     %\textcolor{red}{Saad: plot with number of cases, and show phases are fixed in between valleys}
\end{itemize} 

\subsubsection*{Control Variables}
\added{Our analysis controls for potential confounding factors by incorporating county-level demographic and health variables in our analyses. Specifically, we include the proportion of residents aged 65 and above (sourced from ACS \cite{acs}) to account for age-related COVID-19 risk variations. To control for underlying health conditions known to affect COVID-19 susceptibility and severity we also utilize age-adjusted prevalence of nine key comorbidities \cite{centers2024underlying}: high blood pressure, cancer, diabetes, obesity, stroke, chronic obstructive pulmonary disease (COPD), Chronic kidney disease (CKD), current asthma, and coronary heart disease (CHD) extracted from the CDC PLACES \cite{CDC_PLACES_2022} dataset. These specific health conditions were selected based on extensive epidemiological evidence linking them to increased COVID-19 severity and mortality \cite{yang2020prevalence}.}
%\todo{Saad, can you please add the list of health conditions in the previous sentence?}
%\vanessa{Saad, Reviewer\#3 is also asking for a justification about why these control variables and not others, please add, make sure you also add some references to other papers that use these controls. }

\iffalse
\begin{table}
    \centering
    \begin{tabular}{cc}
        \hline
        Phases & Date Ranges \\
        \hline
        %Phase 0 & March 18, 2020 - June 3, 2020\\
        Phase 0 & June 3, 2020 - September 9, 2020\\
        Phase 1 & September 9, 2020 - March 17, 2021\\
        Phase 2 & March 17, 2021 - June 23, 2021\\
        Phase 3 & June 23, 2021 - November 3, 2021\\
        Phase 4 & November 3, 2021 - March 30, 2022\\
        Phase 5 & March 30, 2022 - June 22, 2022 \\
        Phase 6 & June 22, 2022 - October 20, 2022 \\
        \hline
    \end{tabular}
    \caption{COVID-19 Pandemic Phase Timelines}
    \label{tab:phases}
\end{table}
\fi

\subsection*{Analytical Approach}
\added{To systematically evaluate fairness across race, ethnicity, and urbanization levels in COVID-19 case predictions from the Forecast Hub, we developed a three-step analytical approach that combines error calculation, fairness assessment across sensitive attributes, and interaction analysis with model-data characteristics, while controlling for relevant demographic and health factors.}

%\vanessa{Saad, I would remove this summary because at this stage we have not introduced PBL yet. }
%\added{
%Our analysis focuses on two thrusts: \begin{enumerate}[i.]
%    \item Quantifying the fairness of the COVID-19 case predictions from the Forecast Hub models, with fairness measured as county PBL error parity for two protected attributes: 
%the disparities in COVID-19 case forecast errors with respect to two protected attributes: 
%race or ethnicity and urbanization level \textit{i.e.,} we evaluate whether there exist statistically significant differences between the PBL errors across \raceethnicity  groups as well as across \urbanizationlevel; and 
%\item Quantifying how the relationships between PBL errors and sensitive attributes interact with \texttt{model-data characteristics.}
%(ii) quantifying the changes in fairness when county PBL errors are stratified by \texttt{model-data characteristics.} % including %\added{ \modeltype, training data used (w/o mobility) or \mobilityused, \lookahead  and pandemic \phase.} 
%\end{enumerate}
%To carry out this analysis, we follow \added{two} steps.
%}

\textbf{Step One: Calculating Pinball Loss.} We focus on county error parity as a measure of fairness. %regression fairness. 
Hence, we first need to compute the weekly forecast error at the county level for all the counties in the US. 
To evaluate the accuracy of COVID-19 case forecasts, and given that forecasts in the Forecast Hub are uploaded by teams as quantiles, we employ the pinball loss metric (PBL). This metric quantifies the error of a probabilistic forecast by measuring the distance between observed values and the predicted quantiles, penalizing over- or under-estimation asymmetrically to reflect the actual cost of errors in the prediction. The pinball loss $L_{\tau}(y, f)$ is represented as: $(\tau - 1) \cdot (y - f), \text{if } y < f$ and as $\tau \cdot (f - y),  \text{if } y \geq f$, 
%$$L_{\tau}(y, f) = 
%\begin{cases} 
%(\tau - 1) \cdot (y - f) & \text{if } y < f, \\
%\tau \cdot (f - y) & \text{if } y \geq f,
%\end{cases}$$
where $L_{\tau}(y, f)$ denotes a county's pinball loss for a given quantile $\tau$, $y$ is the observed value \added{i.e., reported number of cases or ground truth extracted from the JHU CSSE COVID-19 case counts dataset \cite{dong2020interactive}}, and $f$ is the forecasted value at quantile $\tau$. For our analysis, we use the average county PBL, computed across the set of $7$ quantiles: $\overline{PBL_{\tau}} = \frac{1}{7} \sum_{i=1}^{7} L_{\tau}(y_i, f_i)$, and normalized by the county population.

%\vanessa{Why is the PBL highlighted like that with $pinball loss norm$? I would just write PBL}
%\saad{Removed.}

%\textbf{Step Two: Relationship between Errors and Race, Ethnicity and Urbanization Levels }
\textbf{Step Two: Fairness across Sensitive Attributes }
%\textbf{Step Two: Relationship between Errors and Sensitive Attributes }
\added{Once weekly average PBLs have been computed per county, we aim to evaluate associations between county prediction errors and the two protected attributes, \raceethnicity and \urbanizationlevel, using regressions. Coefficient analyses will enable us to identify cases in which error parity is violated, pointing to a lack of prediction fairness i.e., significantly different errors across racial or ethnic groups or across urbanization levels.  }

%Once weekly average PBLs have been computed per county, we need to associate counties - and their corresponding prediction errors - with a protected attribute: race or ethnicity as well as urbanization level; these associations will then be used to quantify similarities and differences between errors and attributes - error parity - in Step Three. 
Associating counties, and their PBL errors, to \urbanizationlevel is straight forward using the CDC urban-rural classification scheme \cite{cdcurban} that associates a county with a given urbanization level. 
%\saad{given its more intuitive and interpretable for readers, without essentially losing the urban-rural gradient. In addition to that, it 
%\textcolor{red}{(see Table \ref{X} in the Appendix for specific counts)}. 
On the other hand, associating counties, and their PBL prediction errors, with race and ethnicity would require access to race-stratified predictions. However, due to systemic data collection \added{challenges} during the pandemic, the race-stratified COVID-19 case data necessary to compute race-stratified prediction errors were not collected, hence limiting the predictions provided by the Forecast Hub to county level predictions across all racial and ethnic groups ~\cite{kader2021participatory}. Next, we describe how we proceed in the evaluation of the relationship between forecast errors and sensitive attributes race, ethnicity and urbanization level.

\added{\textbf{Race and Ethnicity Analysis.} To evaluate whether there exist differences between PBL error distributions across racial and ethnic groups, we conduct a regression analysis where county PBL errors are regressed against the racial and ethnic percentage distributions within each county (with White race as the reference group).
%to avoid collinearity). 
In the regression, we control for underlying health conditions, percentage of the population aged 65 and above, the state and data-model characteristics (i.e., type of predictive model, mobility data, lookahead and phase). 
%, health conditions (HO) relevant to COVID-19, and the percentage of the population aged 65 and above. 
An analysis of the resulting regression coefficients can provide insights into how different racial compositions might be positively or negatively associated with forecast accuracy, potentially identifying unfairly higher errors for certain racial or ethnic groups. The main effects regression model, \textbf{Model-1}, can be formally defined as: }

%\vanessa{Saad, Can we use $\sum_{i,j}$ here as well to make it the same as the interaction analysis, instead of two sums?}

\added{
\begin{equation}
PBL_{c,t} = \alpha + \sum_i \beta_i \cdot Race_{i,c} + \sum_{j,k} \gamma_{j,k} \cdot DM_{j,k,c,t} + \sum_l \sigma_l \cdot HO_{l,c} + \theta \cdot age65_{c} + \sum_s \lambda_s \cdot State_{s,c} + \epsilon_{c,t}
\label{eq:glm-1}
\end{equation}}

where:
\begin{itemize}
%    \item \( PBL \): The dependent variable.
    \setlength\itemsep{0.01em}
    \item \( PBL_{c,t} \): The dependent variable representing prediction error for county c at time t.
    \item \( \alpha, \epsilon_{c,t} \): The intercept term and the error terms for county c at time t, respectively.
      \item \( \beta_{\text{Asian}}, \beta_{\text{Black}}, \beta_{\text{Hispanic}} \): Relationship between each race category (Asian, Black, and Hispanic) and \(PBL\), with White as the reference category.
      %, capturing the effect of each race on \( PBL \).
    \item \( \gamma_{jk} \): Relationship between each model-data characteristic $j$ with corresponding category $k$ and the PBL, with $j \in $ \{Lookahead, Phase, Model Type, Mobility Used\}. For example, for the characteristic Lookahead, $k \in (14,21,28)$ (7-days reference group). Reference groups for each model-data characteristic are: 7-days, phase 0, compartmental model and no mobility data used. 
   
    \item \(\sigma_l\): Relationship between age-adjusted health outcomes ($HO$) and the PBL, with $l \in $ \{BPHIGH, CANCER, DIABETES, OBESITY, STROKE, COPD, KIDNEY, CASTHMA, CHD\}. For example, $\sigma_{Asthma}$ represents the effect of asthma prevalence in a given county on the PBL.  
%     \item \( \sum_k \sigma_k \cdot \text{HO}_k \): The combined effects of age-adjusted health outcomes ($HO$) we are considering in the regression. For example, $HO_{Asthma}$ represents the value of the HO, and $\sigma_{Asthma}$ represents the coefficient. 
    \item \( \theta \): Relationship between the percentage of population aged 65 and older in a given county and the \( PBL \).
    \item \( \lambda_s \): Individual state $s$ fixed effect on the error distribution, with $s \in $ \{List of States in US\}.
    %, where \( \lambda_s \) accounts for the effect of individual states on the error distribution. 
    %represents the unique impact of each state.
\end{itemize}
%\noindent where PBL is the dependent variable representing pinball_loss_norm with race being Asian, Black, Hispanic (White as reference), $DM$ representing the \texttt{data-model characteristics}, $HO$ representing the age-adjusted health outcomes after GVIF and $age65$ representing the percentage of population aged $65+$. The State represents the State fixed effects, with $\lambda_s $ representing the impact of state. $\epsilon$ represents the error term }

\added{\textbf{Urbanization Analysis.} For the \urbanizationlevel we replace the Race percentage variables in Equation \ref{eq:glm-1} with the three urbanization levels and we refer to this as \textbf{Model-2}.}

\begin{equation}
PBL_{c,t} = \alpha + \sum_i \beta_i \cdot Urb_{i,c} + \sum_{jk} \gamma_{jk} \cdot DM_{j,k,c,t} + \sum_l \sigma_l \cdot HO_{l,c} + \theta \cdot age65_{c} + \sum_s \lambda_s \cdot State_{s,c} + \epsilon_{c,t}
\label{eq:glm-2}
\end{equation}
\noindent where, $Urb_{i,c}$ where, $Urb_{i,c}$ represents the urbanization level $i$ for county $c$, with $i \in {SMM, MC}$, with LM being the reference category.

\textbf{Step Three: Interaction with Model-Data Characteristics}
\added{Building upon the main effects models, we are also interested in looking into whether the fairness metrics across sensitive attributes change when model and/or data characteristics are considered. In other words, we evaluate if the relationship between county forecast errors and their racial/ethnic and urbanization groups changes when model and data characteristics are taken into account. Next, we describe the methodological approach for each sensitive attribute. }

%\textbf{Step Three: Interaction Analysis between Race, Ethnicity, Urbanization and Model Characteristics}
\added{\textbf{Race and Ethnicity Analysis.} To evaluate whether the relationship between PBL error distributions and race/ethnicity changes across model or data characteristics, we regress the county PBL errors against racial and ethnic percentages for that county (with the White group as a baseline) while adding interaction terms between race/ethnicity and each model-data characteristic, and while controlling for age-adjusted Health Outcome prevalence, the percentage of individuals in the county who are 65+, and the county state. Since we are interested in four model-data characteristics, namely model type, use of mobility data, lookahead and phase, we create for separate regressions (\textbf{Model-1a, -1b, -1c and -1d}, respectively) to account for both main and interaction effects for each model-data characteristic: }

\begin{align*}
PBL_{c,t} = \alpha &+ \sum_i \beta_i \cdot Race_{i,c} + \sum_{jk} \gamma_{j,k} \cdot DM_{j,k,c,t} + \sum_{i,k} \delta_{ijk} \cdot (Race_{i,c} \times DM_{j,k,c,t}) \\
&+ \sum_l \sigma_l \cdot HO_{l,c} + \theta \cdot age\_65_{c} + \sum_s \lambda_s \cdot State_{s,c} + \epsilon_{c,t} 
\tag{3} \label{eq3}
\end{align*}
where
\noindent $\delta_{ijk}$ represents the relationship between the PBL and changes in a given race percentage $i$ and model-data characteristic category $j,k$ with respect to their reference groups (e.g., $\delta_{Hispanic,Lookahead,14-days}$ helps us evaluate how the PBL changes for a 1\% increase in $i=Hispanic$ county population with respect to the White reference group, with $j=Lookahead$ and $k=14-days$ with respect to reference category 7-days). All 
%interaction terms between race/ethnicity and data-model characteristics, with all 
other terms remain the same as Equation \ref{eq:glm-1}.
To be able to evaluate significant changes in the error distribution (PBL) between a minority race and the White reference group for a given model-data characteristic category $j,k$,
we will examine the relative effect computed as
${\beta_i + \delta_{ijk}}$.

%\noindent $\sum_{i,k} \delta_{ijk} \cdot (Race_i \times DM_{jk})$ represents the interaction terms between race/ethnicity and data-model characteristics, with all other terms remaining the same as Equation \ref{eq:glm-1}}. 

%\vanessa{Saad, I have re-written parts of the paragraph above, please see if you agree. I have removed the Exp in the relative effect because the exp is due to the log(DV) that we explain in the next section, right?}

%\added{
%For each of the models 1a - 1d, the data-model characteristic, $DM_{j}$, in the interaction changes each time. The interaction term $\sum_i \sum_k \delta_{ijk} \cdot (Race_i \times DM_{jk})$ captures how racial/ethnic disparities (Asian, Black, Hispanic vs White reference) vary across the categories of each model-data characteristic. Rather than using a single model with all characteristics, we estimate four separate models to isolate how each characteristic interacts with racial/ethnic composition. To understand the full impact of model-data characteristics on racial/ethnic disparities, we examine the exponentiated relative effect ($e^{\beta_i + \delta_{ijk}}$) rather than individual coefficients. This relative effect reveals the disparities with $Race_{i}$ compared to $Race_{white}$ across different data-model characteristics.}

%\vanessa{Saad, correct these formulas to represent the four models we are now considering 1a...1d}
%\vanessa{Saad, add meaning of i=1..3 with reference group..., j=1..4 with each reference group and list all race/eth and DMs again}
%\saad{added.}

\added{\textbf{Urbanization Analysis.} Similarly to the race and ethnicity analysis, we construct four regression models (\textbf{Model-2a,-2b, -2c and -2d}) that mirror the race/ethnicity interaction models, but replace racial percentages with urbanization categories. Using Large Metropolitan (LM) areas as the reference group, we examine interactions between urbanization levels (SMM and MC) and each model-data characteristic. The regression equation takes the form:
\begin{align*}
PBL_{c,t} = \alpha &+ \sum_i \beta_i \cdot Urb_{i,c} + \sum_{jk} \gamma_{jk} \cdot DM_{j,k,c,t} + \sum_l \sigma_l \cdot HO_{l,c} \\
&+ \sum_{i,k} \delta_{ijk} \cdot (Urb_{i,c} \times DM_{j,k,c,t}) + \theta \cdot age\_65_{c} + \sum_s \lambda_s \cdot State_{s,c} + \epsilon_{c,t} \label{eq:eq4}\tag{4}
\end{align*}
where $Urb_{i,c}$ represents the urbanization level $i$ for county $c$,  and the interaction term $\delta_{ijk}$ captures the relationship between a given urbanization level and model-data characteristic category $j,k$ with respect to their reference groups, on the PBL. We will examine the relative effect of a given urbanization level $i$ on the PBL (with respect to LM) and for a given model-data characteristic category $j,k$ computing the relative effect as 
${\beta_i + \delta_{ijk}}$.
}

%the urbanization categories (SMM and MC), and the interaction term $\delta_{ijk}$ captures 
%the relationship of a given urbanization level and model-data characteristic category on the PBL.}

%\vanessa{I have also edited the last sentence above, please check.}
%how urbanization-level disparities vary across different model-data characteristics, while having the same control variables as in the previous models.}

\added{Table \ref{tab:glm_summary} provides a comprehensive overview of our regression framework, detailing the dependent and independent variables for each model. The table systematically presents our main effects models (Model-1 and Model-2) and their corresponding interaction models, along with the control variables and model specifications used throughout our analysis.}

\begin{table*}[h]
\caption{Summary of Generalized Linear Models and Their Specifications}
\label{tab:glm_summary}
\resizebox{\linewidth}{!}{
\begin{tabular}{llll}
\hline
\textbf{Model} & \textbf{Sensitive Attributes} & \textbf{Interaction Terms} & \textbf{Control Variables} \\
\hline
\multicolumn{4}{l}{\textit{Main Effects Models}} \\
\hline
GLM-1 & Race/Ethnicity & None & Health Outcomes \\
      & (\% Black, Hispanic, Asian) & & Data-Model Characteristics \\
      & \% White (ref.) & & State Fixed Effects \\
\hline
GLM-2 & Urbanicity & None & Health Outcomes \\
      & (SMM, MC), & & Data-Model Characteristics \\
      & LM (ref.) & & State Fixed Effects \\
\hline
\multicolumn{4}{l}{\textit{Main + Interaction Effects: Race Interaction Models}} \\
\hline
GLM-1a & Race/Ethnicity & Race × Lookahead & Same as GLM-1 \\
GLM-1b & Race/Ethnicity & Race × Phase & Same as GLM-1 \\
GLM-1c & Race/Ethnicity & Race × Model Type & Same as GLM-1 \\
GLM-1d & Race/Ethnicity & Race × Mobility Used & Same as GLM-1 \\
\hline
\multicolumn{4}{l}{\textit{Main + Interaction Effects: Urbanicity Interaction Models}} \\
\hline
GLM-2a & Urbanicity & Urbanicity × Lookahead & Same as GLM-2 \\
GLM-2b & Urbanicity & Urbanicity × Phase & Same as GLM-2 \\
GLM-2c & Urbanicity & Urbanicity × Model Type & Same as GLM-2 \\
GLM-2d & Urbanicity & Urbanicity × Mobility Used & Same as GLM-2 \\
\hline

\multicolumn{4}{p{0.95\linewidth}}{\scriptsize\textbf{Distribution Family:} Gaussian with log link function for all models} \\
\multicolumn{4}{p{0.95\linewidth}}{\scriptsize\textbf{Health Outcomes:} Asthma, Obesity, COPD, CHD, CKD, Diabetes, Obesity, Cancer, Stroke, Age 65+} \\
\multicolumn{4}{p{0.95\linewidth}}{\scriptsize\textbf{Data-Model Characteristics:} Lookaheads (7(reference) , 14, 21, 28), Phase (0 (reference) 1-6), Model Type (Compartmental (reference), Baseline, Deep Learning, Ensemble, Statistical), Mobility Usage (No (reference), Yes, Mixed)} \\
\multicolumn{4}{p{0.95\linewidth}}{\scriptsize\textbf{Dependent Variable:} Square root of PBL (sqrt\_pbl) with 1\% trimming} \\
\multicolumn{4}{p{0.95\linewidth}}{\scriptsize\textbf{Model Assumptions:}  Independent observations, exponential family distribution (Gaussian), linear predictor through log link function, constant variance of residuals on link scale, no perfect multicollinearity (adjusted GVIF $<$ 2)}  \\
\hline
\end{tabular}}
\end{table*}

\subsection*{Model Selection and Evaluation}
%To control for demographic and health factors at the county level, we incorporate several variables from the American Community Survey (ACS) and PLACES dataset. These include the percentage of residents aged 65 and older, as well as the prevalence of various health conditions: COPD, Chronic Kidney Disease, Obesity, and Chronic Heart Disease. However, 

%, that falls below a threshold of $2$. This process helps us mitigate multicollinearity concerns in our regression models.}

%\vanessa{We need to agree on whether we are going to say PBL os pinball loss; but pinball loss norm might be too much detail?} \saad{PBL is good!}

%\added{Our choice of regression model was guided by careful analysis of the dependent variable's distribution and extensive model diagnostics. As shown in Figure XX in \nameref{s1appendix}, 

%\added{Diagnostic model plots showed substantial heteroscedasticity in residuals, indicating that basic linear regression assumptions were violated (see Section 2.3 in \nameref{s1appendix} for details). In addition, the dependent variable (pinball loss norm) exhibits two key characteristics that influenced our modeling choice for Generalized Linear Models (GLMs): (1) a strong right-skewed distribution with a heavy tail, and (2) strictly positive values (see Figure XX in \nameref{s1appendix}). }

\added{Initial exploratory analysis of the Pinball Loss norm (PBL) distribution revealed several characteristics that influenced our model selection. The original distribution exhibited extreme right-skewness with a heavy tail, through quantile analysis (see Section 2.1 and Fig 2 in \nameref{s1appendix}). These extreme values, while valid measurements, can exert excessive leverage in regression models and potentially obscure patterns in the majority of the data. Therefore, we made the methodological choice to trim the top 1\% values, preserving 99\% of our observations while substantially improving the stability of the model. We also consider squared root transformation during model evaluation.
Given these distributional characteristics - which violate basic linear regression assumptions- and the strictly positive nature of our dependent variable, we tested
both Gamma and Gaussian GLM families with log-link and identity-link functions.}

%Initial exploratory analysis revealed several challenges for standard linear regression. 
\added{ 
Initial analysis revealed substantial multicollinearity among independent variables in the GLM models, particularly between health variables and racial demographics. To address this, we employed an iterative variable selection process using the Generalized Variance Inflation Factor (GVIF\footnote{https://rdrr.io/cran/glmtoolbox/man/gvif.html}) \cite{fox1992generalized}. Variables were retained only if their adjusted GVIF values (calculated as $(\text{GVIF}^{\frac{1}{2 \times \text{Df}}})$, where Df represents degrees of freedom) fell below a threshold of 2. Importantly, we proceeded with our analysis only after confirming that our variables of interest - race, ethnicity, urbanization level and data-model characteristics - all demonstrated adjusted GVIF values below this threshold, ensuring the reliability of our primary coefficient estimates.
}

%combined with four different data combinations (no transformation (all),  square root (all), no transformation (trimmed) and square root (trimmed)). 
\added{Using diagnostic plots, residual patterns, and model performance metrics (pseudo-$R^2$ values) for model evaluation, we identified the Gaussian GLM with log-link function applied to square root transformed data (with 1\% trimming) as the best fit. Multicollinearity analyses and model diagnostics for these GLMs are discussed in depth in Sections 2.2 and 2.3 in \nameref{s1appendix}. }
%\vanessa{Add appendix section or figure numbers.} \saad{Will do, once I reorganize the paper first, and then the appendix, or else the numbers will be jumbled.}

%\textcolor{red}{IMPORTANT: Saad, did you compute the square root of all variables (DV and IV) or only the DV). I am asking because when we discuss results, if only DV is squared root then we need to either change discussion to say " 1\% increase in Hispanic pop was associated with a 2\% increase in the SQUARED ROOT PBL error, instead of the PBL error (or maybe clarify at the beginning of each discussion that when we say PBL we mean squared root). If the square root was applied to all variables, then I think we should be fine. Please let me know your thoughts. }

\added{In the next section, we evaluate the relationship of race, ethnicity and urbanization levels with the forecast error (PBL) distribution via Model-1 and Model-2, which we rename to GLM-1 and GLM-2. We also evaluate how that relationship is modulated when interaction terms between race, ethnicity, urbanization levels and the four model and data characteristics are incorporated into the model i.e., Model-1a through Model-1d and Model-2a through Model-2d that we rename as GLM-1a through GLM-1d and GLM-2a through GLM-2d (see Table \ref{tab:glm_summary} for model summary).
Following the Results section, we will present hypotheses for the reasons behind these findings in the Discussion section. 
}

\section*{Results}
\subsection*{Fairness of COVID-19 case predictions across race and ethnicity (GLM-1)}
\added{
Based on Equation~\ref{eq:glm-1}, the generalized linear model (GLM-1) characteristics are illustrated in Table \ref{tab:glm-1}. The table provides insights into the relationship between race, ethnicity and the PBL while controlling for the impact of various data-model features, health outcomes and state fixed effects on prediction accuracy. As a note, some of the health outcome covariates were left out due to inflated GVIF.  }

%\begin{figure}[t!]
%\centering
%\subfloat[Normalized PBL distributions stratified by \\ racial and ethnic groups. \label{fig:pbl_race}]{\includegraphics[width=0.5\textwidth]{figs/Race_group.png}}
%\subfloat[PBL and race: regression approach with coefficient significance (*** $p<0.001$) and 95\% confidence intervals. \label{fig:pbl_race_regression}]{\includegraphics[width=0.5\textwidth]{figs/regr_race.png}}
%\subfloat[Pairwise Mean Differences Between Races and Ethnicities.\label{fig:2a}]{\includegraphics[width=0.45\textwidth]{figs/_tukey_Race.png}}
%\includegraphics[width=1\linewidth]{new_figs/level2RACE/Gamma/log/trimmed/transformed/regression_level2RACE_race_trimmed_tr_sqrt_pbl_Gamma_log_coef_plot_race_and_model.png}

%\caption{Forest plot demonstrating the distribution of the exponential coefficients for GLM-1. Noting that the White Race, Phase-1, Lookahead-7 days and Mobility Used - No are the reference groups. The State-Fixed effect coefficients are left out.
%}
%\label{fig:fp_glm1}
%\end{figure}

% Required package for table formatting

\begin{table}[htbp]
\caption{ Summary of Regression Results: GLM-1}
\label{tab:glm-1}
\scriptsize
\resizebox{\linewidth}{!}{\begin{tabular}{lcccc}
\hline
\textbf{Variable} & \textbf{$\mathbf{e^{Coef.}}$ (Std. Error)} & \textbf{95\% CI} & \textbf{z-value} \\
\hline
Intercept & 0.009$^{***}$ (0.009) & [0.009, 0.009] & -510.099 \\
\rowcolor[gray]{0.9} & \multicolumn{3}{l}{\textit{Sensitive Attributes}} \\
\rowcolor[gray]{0.9}\% Black & 0.976$^{***}$ (0.003) & [0.970, 0.983] & -6.864 \\
\rowcolor[gray]{0.9}\% Hispanic & 1.216$^{***}$ (0.003) & [1.209, 1.223] & 63.844 \\
\rowcolor[gray]{0.9}\% Asian & 0.515$^{***}$ (0.015) & [0.500, 0.530] & -44.647 \\
& \multicolumn{3}{l}{\textit{Lookahead Period}} \\
14 days & 1.119$^{***}$ (0.001) & [1.118, 1.121] & 143.586 \\
21 days & 1.211$^{***}$ (0.001) & [1.209, 1.213] & 249.396 \\
28 days & 1.300$^{***}$ (0.001) & [1.298, 1.302] & 352.668 \\
& \multicolumn{3}{l}{\textit{Phase Effects}} \\
Phase 1 & 1.445$^{***}$ (0.001) & [1.443, 1.448] & 377.204 \\
Phase 2 & 0.890$^{***}$ (0.001) & [0.888, 0.892] & -99.535 \\
Phase 3 & 1.285$^{***}$ (0.001) & [1.283, 1.288] & 236.425 \\
Phase 4 & 1.553$^{***}$ (0.001) & [1.549, 1.556] & 416.701 \\
Phase 5 & 0.969$^{***}$ (0.001) & [0.967, 0.972] & -22.748 \\
Phase 6 & 1.152$^{***}$ (0.001) & [1.149, 1.154] & 112.373 \\
& \multicolumn{3}{l}{\textit{Model Type}} \\
Baseline & 1.121$^{***}$ (0.002) & [1.118, 1.124] & 72.937 \\
Deep Learning & 1.037$^{***}$ (0.001) & [1.035, 1.039] & 40.193 \\
Ensemble & 1.007$^{***}$ (0.001) & [1.005, 1.009] & 7.959 \\
Statistical & 1.076$^{***}$ (0.001) & [1.074, 1.078] & 89.646 \\
& \multicolumn{3}{l}{\textit{Mobility}} \\
Mixed & 0.864$^{***}$ (0.001) & [0.862, 0.866] & -124.255 \\
Yes & 0.996$^{***}$ (0.001) & [0.994, 0.997] & -5.637 \\
& \multicolumn{3}{l}{\textit{Health Controls}} \\
Asthma & 0.959$^{***}$ (0.001) & [0.957, 0.961] & -43.776 \\
Obesity & 1.003$^{***}$ (0.000) & [1.003, 1.004] & 25.942 \\
COPD & 1.046$^{***}$ (0.000) & [1.045, 1.047] & 93.094 \\
\% Age 65+ & 1.214$^{***}$ (0.007) & [1.198, 1.230] & 28.712 \\
\hline
\multicolumn{4}{l}{\footnotesize \textbf{Model Statistics:} Pseudo $R^2$ (Cox-Snell) = 0.460; Log-Likelihood = 5,981,789; N = 1,526,869} \\
\hline
\multicolumn{4}{p{0.95\textwidth}}{\scriptsize \textbf{Notes:} $^{***}p<0.001$, $^{**}p<0.01$, $^{*}p<0.05$. \textit{Dependent Variable:} sqrt\_pbl. \textit{Link Function:} log. \textit{Regression Family: } Gaussian. State fixed effects included but are reported in Table 5 in \nameref{s1appendix}. \textit{exp(Coef)} represents the multiplicative effect on the outcome. CI: Confidence Interval.} \\
\hline
\end{tabular}}
\end{table}

%\subsection{Does race and rurality have an impact on COVID-19 case prediction errors?}
%\subsection{Analysis of COVID-19 case predictions PBLs across race groups}

\added{For the race and ethnicity variables, the regression results indicate that
for every 1\% increase in a county's Hispanic population (with respect to the reference group White) the prediction errors increased by approximately $(1.216 - 1) * 100 \% = 21.6\%$
%the percentage of Hispanic population in a county was associated with significantly higher prediction errors 
($\beta$ = 0.196, $\exp(\beta)$ = 1.216, $p < 0.001$).
%, suggesting that our models were approximately 21.6\% less accurate in areas with higher Hispanic populations. 
In contrast, regions with larger Asian populations showed markedly lower prediction errors than the White population baseline ($\beta = -0.663$, $\exp(\beta) = 0.515$, $p < 0.001$). Our results indicate that for every 1\% increase in the Asian population (with respect to White population), the prediction accuracy is  
%indicating that the predictions were more accurate in these areas. The model performed 
approximately 48.5\% better when compared to the White group. 
For predominantly Black communities, we observed a slight but statistically significant improvement in prediction accuracy ($\beta = -0.024$, $\exp(\beta) = 0.976$, $p < 0.001$), with 1\% increase in Black population (with respect to White) being associated with predictions approximately 2.4\% more accurate than the baseline.}

%\added{Interestingly, all of the health outcome control variables (except asthma) showed a significant positive relationship with the COVID-19 prediction errors, with increases in 1\% prevalence of obesity, COPD or age 65+ associated with increases in the prediction errors of 0.3\%, 4.6\% and 21.4\%, respectively.}

\added{\textbf{Model diagnostics.} Model assessments  indicate that Gaussian-GLM demonstrates adequate fit and reliability. The pseudo-$R^2$ is 0.46, and the residuals show an approximately normal distribution with mild heteroskedasticity at extreme values. Cook's distance analysis identified no influential points that would substantially affect our findings. Although there are some deviations from normality in the tails of the residual distribution, these do not materially affect our main conclusions regarding demographic disparities in prediction accuracy (detailed diagnostics are provided in Section 2.3.1, Fig 3 in \nameref{s1appendix}). }

%\vanessa{table 1 says Tsq is 0.46 but the text says 0.38. please correct.}

\begin{tcolorbox}[title=\textbf{Summary}: Race, colback=gray!10, colframe=gray!50]

COVID-19 forecast errors show substantial racial disparities, particularly for \textbf{Hispanic} Race:
\begin{itemize}
    \setlength\itemsep{0.1em}
    \renewcommand{\labelitemi}{$\Rightarrow$}    
    \item 21.6\% \textit{higher} for Hispanic populations
    \item 48.5\% \textit{lower} for Asian populations
    \item 2.4\% \textit{lower} for Black population
\end{itemize} %48.5% lower for Asian populations, and 2.4% lower for Black populations compared to White communities, per 1% population increase.%In the Discussion section, we discuss potential reasons behind these findings.

%This calls for a further investigation into both the methodological and structural factors that can contribute to these prediction inequities. }
\end{tcolorbox}

\subsection*{Fairness of COVID-19 case predictions across urbanization levels (GLM-2)}
\added{
Table \ref{tab:glm2} illustrates the estimated coefficients of the generalized linear model for the fairness analysis of COVID-19 prediction accuracy across urbanization levels (GLM-2, Equation \ref{eq:glm-2}). The table reveals insights into the relationship between urbanization levels and the PBL while controlling for the impact of various data-model features, health outcomes and state fixed effects on prediction accuracy. }

\begin{table}[h]
\caption{Summary of Regression Results: GLM-2}
\label{tab:glm2}
\scriptsize
\resizebox{\linewidth}{!}{\begin{tabular}{lcccc}
\hline
\textbf{Variable} & \textbf{$\mathbf{e^{Coef.}}$ (Std. Error)} & \textbf{95\% CI} & \textbf{z-value} \\
\hline
Intercept & 0.009$^{***}$ (0.018) & [0.009, 0.009] & -263.017 \\
\rowcolor[gray]{0.9} & \multicolumn{3}{l}{\textit{Sensitive Attributes}} \\
\rowcolor[gray]{0.9}Micropolitan & 1.065$^{***}$ (0.001) & [1.063, 1.067] & 60.326 \\
\rowcolor[gray]{0.9}Small and Medium Metro & 1.027$^{***}$ (0.001) & [1.025, 1.029] & 26.509 \\
& \multicolumn{3}{l}{\textit{Lookahead Period}} \\
14 days & 1.119$^{***}$ (0.001) & [1.117, 1.121] & 143.691 \\
21 days & 1.211$^{***}$ (0.001) & [1.209, 1.213] & 249.533 \\
28 days & 1.299$^{***}$ (0.001) & [1.298, 1.301] & 352.841 \\
& \multicolumn{3}{l}{\textit{Phase Effects}} \\
Phase 1 & 1.445$^{***}$ (0.001) & [1.442, 1.448] & 377.540 \\
Phase 2 & 0.890$^{***}$ (0.001) & [0.888, 0.892] & -100.050 \\
Phase 3 & 1.285$^{***}$ (0.001) & [1.282, 1.287] & 236.167 \\
Phase 4 & 1.552$^{***}$ (0.001) & [1.549, 1.555] & 416.754 \\
Phase 5 & 0.969$^{***}$ (0.001) & [0.967, 0.972] & -22.827 \\
Phase 6 & 1.151$^{***}$ (0.001) & [1.148, 1.154] & 112.311 \\
& \multicolumn{3}{l}{\textit{Model Type}} \\
Baseline & 1.121$^{***}$ (0.002) & [1.118, 1.124] & 73.043 \\
Deep Learning & 1.037$^{***}$ (0.001) & [1.035, 1.039] & 40.217 \\
Ensemble & 1.007$^{***}$ (0.001) & [1.005, 1.009] & 7.952 \\
Statistical & 1.076$^{***}$ (0.001) & [1.074, 1.078] & 89.919 \\
& \multicolumn{3}{l}{\textit{Mobility}} \\
Mixed & 0.864$^{***}$ (0.001) & [0.862, 0.866] & -124.509 \\
Yes & 0.996$^{***}$ (0.001) & [0.994, 0.997] & -5.631 \\
& \multicolumn{3}{l}{\textit{Health Controls}} \\
Asthma & 0.956$^{***}$ (0.001) & [0.954, 0.958] & -34.853 \\
Obesity & 1.004$^{***}$ (0.000) & [1.004, 1.005] & 32.018 \\
BPHigh & 0.998$^{***}$ (0.000) & [0.997, 0.998] & -7.667 \\
COPD & 1.000 (0.001) & [0.998, 1.002] & -0.089 \\
Stroke & 0.981$^{***}$ (0.003) & [0.974, 0.988] & -5.605 \\
Cancer & 0.984$^{***}$ (0.003) & [0.979, 0.990] & -5.650 \\
CHD & 1.089$^{***}$ (0.003) & [1.084, 1.095] & 33.418 \\
Diabetes & 0.995$^{***}$ (0.001) & [0.993, 0.997] & -5.442 \\
CKD & 1.043$^{***}$ (0.006) & [1.030, 1.056] & 6.559 \\
\% Age 65+ & 1.111$^{***}$ (0.007) & [1.096, 1.126] & 15.328 \\
\hline
\multicolumn{4}{l}{\scriptsize \textbf{Model Statistics:} Pseudo $R^2$ (CS) = 0.462; Log-Likelihood = 5,983,292; N = 1,526,869} \\
\hline
\multicolumn{4}{p{0.95\textwidth}}{\scriptsize \textbf{Notes:} $^{***}p<0.001$, $^{**}p<0.01$, $^{*}p<0.05$. \textit{Dependent Variable:} sqrt\_pbl. \textit{Link Function:} log. \textit{Regression Family:} Gaussian. State fixed effects included but are reported in Table 6 in \nameref{s1appendix}. $exp(Coef)$ represents the multiplicative effect on the outcome. CI: Confidence Interval.} \\
\hline
\end{tabular}}
\centering
\end{table}

%\begin{figure}[!t]
%\includegraphics[width=1\linewidth]{new_figs/regression_level2URBANICITY_race_trimmed_tr_sqrt_pbl_Gaussian_log_coef_plot_urban_and_model.png} 
%\caption{Forest plot demonstrating the distribution of the exponential coefficients for GLM-1b. Noting that the Large Metropolitans(LM), Phase-1, Lookahead-7 days, Model Type-Compartmental and Mobility Used - No are the reference groups. The State-Fixed effect coefficients are left out.}
%\label{fig:fpglm2}
%\end{figure}

\added{Relative to LM areas, prediction errors varied significantly across urbanization levels. MC areas, which are urban clusters with populations up to 50,000, showed notably higher prediction errors ($\beta = 0.063$, $\exp(\beta) = 1.065$, $p < 0.001$), indicating approximately 6.5\% higher errors in these regions. This suggests particular challenges in forecasting COVID-19 cases in smaller urban areas that may have different healthcare infrastructure and reporting systems compared to LM areas. SMM areas also demonstrated increased prediction errors ($\beta = 0.027$, $\exp(\beta) = 1.027$, $p < 0.001$), though the effect was more modest with 2.7\% higher errors than large metropolitan regions. } %The gradient in error rates across urbanization levels suggests that prediction accuracy may be influenced by factors such as healthcare capacity, testing infrastructure, and reporting mechanisms that vary systematically with urbanization level.}

\added{\textbf{Model diagnostics.} Our model assessments  reveal that GLM-2 achieves similar statistical properties to GLM-1, with a pseudo-$R^2$ of $0.46$. While the residual distribution shows some deviation from normality at the tails, these departures do not substantially impact our key findings regarding urbanicity-based disparities in prediction accuracy. 
The complete characteristics of the model and the GVIF analysis can be found in Section 2.3.2, Fig 4 in \nameref{s1appendix}.}

\begin{tcolorbox}[title=\textbf{Summary}: Urbanicity, colback=gray!10, colframe=gray!50]
%\added{
\begin{itemize}
    \setlength\itemsep{0.0001em}
    \renewcommand{\labelitemi}{$\Rightarrow$}    
    \item Prediction disparities worsen for more rural areas.
    \end{itemize}

%Our analyses reveal significant disparities in COVID-19 forecasting accuracy across urbanization levels. MC areas show notably higher prediction errors (6.5\%) compared to LM areas, while small and medium metropolitan areas demonstrate moderately increased errors (2.7\%). 
%}
\end{tcolorbox}

\subsection*{Fairness of COVID-19 case predictions across race, ethnicity and model-data characteristics}

\subsubsection*{GLM-1a: Race, Ethnicity and Forecast Lookahead}
\added{
%Next, using coefficients from regression found from Equation \ref{eq:eq3} we examined how these racial and ethnic disparities vary across different forecast horizons (lookahead periods). 
Table \ref{tab:glm-1a} presents the interaction effects between race, ethnicity and lookahead periods, revealing notable variations in forecast accuracy across different time horizons, and pointing to unequal (unfair) error distributions (see Equation \ref{eq3}). The trends described for the main effects (GLM-1) persist when adding the interaction effects \textit{i.e.,} increases in Black or Asian population in a county (with respect to White) and for a given lookahead, are associated with lower PBLs; while increases in Hispanic population are associated with higher PBLs.}

%% Lookahead X Race
\begin{table*}[h]
\caption{\textbf{GLM-1a}: Race × Lookahead Effects Relative to White Reference Group}
\label{tab:glm-1a}
\resizebox{\linewidth}{!}{\begin{tabular}{lcc|cc}
\hline
& \multicolumn{2}{c|}{\textbf{Coefficient Estimates}} & \multicolumn{2}{c}{\textbf{Relative Effect}} \\
\hline
\textbf{Variable} & \textbf{e\textsuperscript{Coef.} (SE)} & \textbf{95\% CI} & \textbf{e\textsuperscript{Coef.}} & \textbf{\% Diff from White} \\
\hline
\% Asian (7-day ref.) & 0.228\textsuperscript{***} (0.030) & [0.215, 0.243] & 0.228\textsuperscript{***} & -77.2\% \\
\quad × 14-day ahead & 1.652\textsuperscript{***} (0.037) & [1.536, 1.777] & 0.377\textsuperscript{***} & -62.3\% \\
\quad × 21-day ahead & 2.689\textsuperscript{***} (0.035) & [2.509, 2.882] & 0.613\textsuperscript{***} & -38.7\% \\
\quad × 28-day ahead & 3.505\textsuperscript{***} (0.034) & [3.279, 3.746] & 0.799\textsuperscript{***} & -20.1\% \\

\hline
\% Black (7-day ref.) & 0.937\textsuperscript{***} (0.005) & [0.928, 0.947] & 0.937\textsuperscript{***} & -6.3\% \\
\quad × 14-day ahead & 1.034\textsuperscript{***} (0.006) & [1.023, 1.046] & 0.969\textsuperscript{***} & -3.1\% \\
\quad × 21-day ahead & 1.053\textsuperscript{***} (0.005) & [1.042, 1.065] & 0.987\textsuperscript{***} & -1.3\% \\
\quad × 28-day ahead & 1.067\textsuperscript{***} (0.005) & [1.056, 1.078] & 1.000 & 0.0\% \\
\hline
\% Hispanic (7-day ref.) & 1.298\textsuperscript{***} (0.005) & [1.287, 1.310] & 1.298\textsuperscript{***} & +29.8\% \\
\quad × 14-day ahead & 0.960\textsuperscript{***} (0.005) & [0.951, 0.970] & 1.246\textsuperscript{***} & +24.6\% \\
\quad × 21-day ahead & 0.913\textsuperscript{***} (0.005) & [0.904, 0.923] & 1.185\textsuperscript{***} & +18.5\% \\
\quad × 28-day ahead & 0.900\textsuperscript{***} (0.005) & [0.891, 0.909] & 1.168\textsuperscript{***} & +16.8\% \\
\hline

\end{tabular}}

{\scriptsize \textbf{Model Statistics:} Pseudo $R^2$ (CS) = 0.461; Log-Likelihood = 5,982,892; N = 1,526,869}

{\scriptsize \textbf{Notes:}  \textsuperscript{***}$p<0.001$, \textsuperscript{**}$p<0.01$, \textsuperscript{*}$p<0.05$. \textit{Dependent Variable:} Square root PBL. \textit{Link Function:} Log. \textit{Regression Family:} Gaussian. The table shows the GLM coefficients and their significance for model GLM-1a. We only discuss race, ethnicity and lookahead interaction coefficients. For clarity purposes,  all other main effects and control variables: Health outcomes, age 65+ and state fixed effects are only shown and discussed in Tables 7, 8 in  \nameref{s1appendix}. Model Diagnostics are provided in Fig 5a in \nameref{s1appendix}.
For the Coefficient Estimates, \textit{exp(Coef)} represents the multiplicative effect on the outcome,
%\vanessa{what does the "multiplicated effect on the outcome" mean? what is the outcome, the forecast error PBL? also, do you mean "Coefficient Estimates? it's the same as Coef, right? please clarify}
SE: Standard Error and CI: Confidence Interval. 
The Relative Effect represents the multiplicative effect on the forecast error (PBL) of a particular race or ethnicity compared to the White population within each lookahead. 
%data-model characteristic. 
%\vanessa{where is the marginal effect in the table? is it relative effect?.} \saad{Yes. What we are reporting is relative effect, marginal effect is something different.}
The relative effect is represented by \textit{exp(Coef)} and computed as $e^{\beta_{i} + \delta_{ij}}$ with coefficients from Equation \ref{eq3}.
%, \vanessa{Saad, please add formula}. \saad{added}
To evaluate better the relative effect, we also discuss the percentage change in forecast error when compared to White population for each lookahead variable (\textit{\% Diff from White}). 
This change is computed as $(1- e^{\beta_{i} + \delta_{ij}})*100\%$  
%\vanessa{Saad, add formula, I think it's in the text when you discuss results, please move it from there.} \saad{added}
%The \textit{\% Diff from White}
for a given race/ethnicity and lookahead value, and it represents the percentage increase or decrease in the forecast error (PBL) with respect to the White population (e.g., +24.6\% means that the PBL error for Hispanic counties at 14-day lookahead is 24.6\% higher PBL when compared to White). All effects should be interpreted as the relative difference compared to White population within each specific lookahead value. The relative coefficient significance is evaluated using \texttt{linearHypothesis} in R (\textit{car} package \cite{fox2018r}) } 
\vspace{0.2cm}
\end{table*}

%For counties with higher Black population, while the baseline shows slightly better prediction accuracy compared to White areas ($exp(\beta)$ = 0.937, $p < 0.001$), it is to be noted that the main effects coefficients alone are not itself interpretable, since there are interaction terms. 
%Rather, we need to look at the combined effect ($\beta_{i} + \delta_{ij}$ from Equation \ref{eq:eq3}) to understand the disparities at different forecast horizons. 

\added{Our analysis reveals that increases in Hispanic population (with respect to White) are related to persistent disparities in forecast accuracy across all prediction horizons, though these disparities show a gradual decrease for predictions over longer timeframes.  }
%The base coefficient ($\exp(\beta)$ = 1.298) combined with the interaction terms demonstrates that while the magnitude of disparity reduces with extended forecasting periods (interaction coefficients of 0.960, 0.913, and 0.900 for 14-, 21-, and 28-day horizons respectively, all statistically significant at p < 0.001), substantial differences in prediction accuracy remain. 
\added{When examining the relative effects, we find that areas with higher Hispanic populations experience prediction errors that are 24.6\% higher than predominantly White areas at 14-day forecasts, declining to 18.5\% at 21-day forecasts, and further reducing to 16.8\% at 28-day forecasts. This pattern indicates that although the disparity in prediction accuracy moderates somewhat over longer forecast horizons, significant inequities in model performance persist throughout all prediction timeframes for Hispanic communities.}

\added{For Black counties, the relative effect shows progressively decreasing significant differences in forecast errors between Black and the White reference group, with PBL errors being $3.1\%$ higher for White counties for the 14-day forecast ($exp(\beta)$ = 0.969, $p<0.001$), decreasing to $1.3\% $ higher for 21 days ($exp(\beta)$ = 0.987, $p<0.001$). By the 28-day forecast horizon, the prediction accuracy for Black counties shows no significant difference from White areas (relative effect: 1.000, 0.0\% difference).}
%\added{For Black counties, the interaction terms show 
%progressively increasing coefficients (14-day: $exp(\beta)$ = 1.034; 21-day: $exp(\beta)$ = 1.053; 28-day: $exp(\beta)$ = 1.067; all $p < 0.001$), resulting in relative effect that approach parity with White areas. }

\added{Asian counties reveal the most 
%For counties with higher Asian population, examining the combined effects reveals the most 
pronounced variation across forecast horizons. The relative effect analysis shows that 
%When we combine the main effect ($\exp(\beta)$ = 0.228) with the interaction coefficients (14-day: $\exp(\beta)$ = 1.652; 21-day: $\exp(\beta)$ = 2.689; 28-day: $\exp(\beta)$ = 3.505; all $p < 0.001$), we find that 
while predictions remain more accurate than for White counties across all horizons, this advantage decreases substantially over longer forecast periods. The relative effect shows that areas with higher Asian populations have prediction errors 62.3\% lower than White areas at 14-day forecasts, improving to 38.7\% lower at 21-day forecasts, and 20.1\% lower at 28-day forecasts. }

\subsubsection*{GLM-1b: Race, Ethnicity and COVID-19 Phase}

%\vanessa{Saad, make sure the caption in all tables has all the text in Table 4 (adapted to each model-data char.}

\added{
When examining how racial and ethnic disparities vary across different pandemic phases, we observe substantial heterogeneity in PBL, with patterns varying markedly by both race and phase (see Table \ref{tab:glm-1b}). The results reveal complex temporal dynamics in prediction fairness across different demographic groups.}

%% Race X Phase
\begin{table*}[!ht]

\caption{\textbf{GLM-1b}: Race × Phase Effects Relative to White Reference Group}
\label{tab:glm-1b}
\resizebox{\linewidth}{!}{\begin{tabular}{lcc|cc}
\hline
& \multicolumn{2}{c|}{\textbf{Coefficient Estimates}} & \multicolumn{2}{c}{\textbf{Relative Effect}} \\
\textbf{Variable} & \textbf{e\textsuperscript{Coef.} (SE)} & \textbf{95\% CI} & \textbf{e\textsuperscript{Coef.}} & \textbf{\% Diff from White} \\
\hline
\% Asian (Phase 0 ref.) & 0.137\textsuperscript{***} (0.042) & [0.126, 0.149] & 0.137\textsuperscript{***} & -86.3\% \\
\quad × Phase 1 & 2.092\textsuperscript{***} (0.046) & [1.912, 2.290] & 0.287\textsuperscript{***} & -71.3\% \\
\quad × Phase 2 & 8.767\textsuperscript{***} (0.049) & [7.957, 9.659] & 1.201\textsuperscript{***} & +20.1\% \\
\quad × Phase 3 & 0.812\textsuperscript{***} (0.053) & [0.732, 0.900] & 0.111\textsuperscript{***} & -88.9\% \\
\quad × Phase 4 & 6.310\textsuperscript{***} (0.047) & [5.756, 6.918] & 0.864\textsuperscript{***} & -13.6\% \\
\quad × Phase 5 & 35.577\textsuperscript{***} (0.049) & [32.323, 39.159] & 4.871\textsuperscript{***} & +387.1\% \\
\quad × Phase 6 & 3.260\textsuperscript{***} (0.057) & [2.916, 3.645] & 0.447\textsuperscript{***} & -55.3\% \\
\hline
\% Black (Phase 0 ref.) & 2.130\textsuperscript{***} (0.005) & [2.108, 2.153] & 2.130\textsuperscript{***} & 113.0\% \\
\quad × Phase 1 & 0.438\textsuperscript{***} (0.006) & [0.434, 0.443] & 0.933\textsuperscript{***} & -6.7\% \\
\quad × Phase 2 & 0.393\textsuperscript{***} (0.007) & [0.388, 0.399] & 0.837\textsuperscript{***} & -16.3\% \\
\quad × Phase 3 & 0.541\textsuperscript{***} (0.006) & [0.535, 0.548] & 1.152\textsuperscript{***} & +15.2\% \\
\quad × Phase 4 & 0.270\textsuperscript{***} (0.007) & [0.267, 0.274] & 0.575\textsuperscript{***} & -42.5\% \\
\quad × Phase 5 & 0.375\textsuperscript{***} (0.009) & [0.369, 0.382] & 0.799\textsuperscript{***} & -20.1\% \\
\quad × Phase 6 & 0.529\textsuperscript{***} (0.008) & [0.521, 0.536] & 1.127\textsuperscript{***} & +12.7\% \\
\hline
\% Hispanic (Phase 0 ref.) & 2.018\textsuperscript{***} (0.005) & [1.998, 2.039] & 2.018\textsuperscript{***} & 101.8\% \\
\quad × Phase 1 & 0.677\textsuperscript{***} (0.006) & [0.670, 0.685] & 1.366\textsuperscript{***} & +36.6\% \\
\quad × Phase 2 & 0.549\textsuperscript{***} (0.007) & [0.541, 0.557] & 1.108\textsuperscript{***} & +10.8\% \\
\quad × Phase 3 & 0.573\textsuperscript{***} (0.006) & [0.566, 0.580] & 1.156\textsuperscript{***} & +15.6\% \\
\quad × Phase 4 & 0.455\textsuperscript{***} (0.006) & [0.449, 0.460] & 0.918\textsuperscript{***} & -8.2\% \\
\quad × Phase 5 & 0.494\textsuperscript{***} (0.009) & [0.485, 0.503] & 0.997 & -0.3\% \\
\quad × Phase 6 & 0.555\textsuperscript{***} (0.008) & [0.547, 0.564] & 1.120\textsuperscript{***} & +12.0\% \\
\hline

\end{tabular}}

{\scriptsize \textbf{Model Statistics:} Pseudo $R^2$ (CS) = 0.495; Log-Likelihood = 6,012,390; N = 1,526,869} 

{\scriptsize \textbf{Notes:} \textsuperscript{***}$p<0.001$, \textsuperscript{**}$p<0.01$, \textsuperscript{*}$p<0.05$. \textit{Dependent Variable:} Square root PBL. \textit{Link Function:} Log. \textit{Regression Family:} Gaussian. The table shows the GLM coefficients and their significance for model GLM-1b. We only discuss race, ethnicity and phase interaction coefficients. For clarity purposes, all other main effects and control variables: Health outcomes, age 65+ and state fixed effects are only shown and discussed in Tables 9, 10 in  \nameref{s1appendix}. Model Diagnostics are provided in Fig 5b in \nameref{s1appendix}.
For the Coefficient Estimates, \textit{exp(Coef)} represents the multiplicative effect on the outcome,
SE: Standard Error and CI: Confidence Interval.
The Relative Effect represents the multiplicative effect on the forecast error (PBL) of a particular race or ethnicity compared to the White population within each phase.
The relative effect is represented by \textit{exp(Coef)} and computed as $e^{\beta_{i} + \delta_{ij}}$ with coefficients from Equation \ref{eq3}.
To evaluate better the relative effect, we also discuss the percentage change in forecast error when compared to White population for each phase variable (\textit{\% Diff from White}).
This change is computed as $(1- e^{\beta_{i} + \delta_{ij}})*100\%$
for a given race/ethnicity and phase value, and it represents the percentage increase or decrease in the forecast error (PBL) with respect to the White population (e.g., +101.8\% means that the PBL error for Hispanic counties in Phase 0 is 101.8\% higher PBL when compared to White). All effects should be interpreted as the relative difference compared to White population within each specific phase value. The relative coefficient significance is evaluated using \texttt{linearHypothesis} in R (\textit{car} package \cite{fox2018r}) }
\end{table*}

\added{For Hispanic populations, the analysis reveals significant variation in prediction errors across pandemic phases relative to White populations. At initial stages of the pandemic (Phase 0), Hispanic counties show substantially higher errors (+101.8\%) compared to White counties. While this disparity persists across most phases, its magnitude fluctuates notably. Phase 1 also exhibits a higher disparity (+36.6\%), followed by Phase 3 (+15.6\%) and Phase 6 (+12.0\%). However, in Phases 4 and 5, this pattern reverses, with Hispanic counties showing slightly lower errors than White counties (-8.2\% and -0.3\% respectively), suggesting that prediction fairness for Hispanic communities varied significantly with pandemic phase characteristics. }

\added{For Black populations, the results show complex phase-dependent patterns. Starting with notably higher errors in Phase 0 (+113.0\% compared to White areas), the disparities shift dramatically across phases. Phase 4 shows the most favorable performance for Black communities, with errors 42.5\% lower than White areas. However, Phases 3 and 6 show notably higher prediction errors (+15.2\% and +12.7\% respectively) compared to White areas. Phases 1, 2, and 5 show better performance for Black communities (-6.7\%, -16.3\%, and -20.1\% respectively).}

\added{Asian populations exhibit the most extreme phase-dependent variations in prediction accuracy. While starting with substantially lower errors than White areas in Phase 0 (-86.3\%), the disparities show dramatic swings across phases. Phase 5 stands out with strikingly higher errors (+387.1\% compared to White areas), while Phase 3 shows the best relative performance (-88.9\%). Phase 2 also shows notably higher errors (+20.1\%), while Phases 1, 4, and 6 maintain lower errors compared to White areas (-71.3\%, -13.6\%, and -55.3\% respectively). These extreme variations suggest that predictions for Asian communities were particularly sensitive to phase-specific characteristics of the pandemic.}

\added{All these phase-dependent variations are statistically significant $(p<0.001)$, and they reveal that prediction fairness across racial and ethnic groups is not consistent throughout the pandemic, with certain phases (particularly Phase 0) associated with the largest disparities relative to White populations.} %This suggests that model performance and fairness may be influenced by phase-specific factors such as case volumes, testing patterns, or reporting practices that varied across different racial and ethnic communities during different pandemic phases.}

\subsubsection*{GLM-1c: Race, Ethnicity and Model Type}

\begin{table*}[h]
\caption{GLM-1c: Race × Model Type Effects Relative to White Reference Group}
\label{tab:glm-1c}
\resizebox{\linewidth}{!}{\begin{tabular}{lcc|cc}
\hline
& \multicolumn{2}{c|}{\textbf{Coefficient Estimates}} & \multicolumn{2}{c}{\textbf{Relative Effect}} \\
\textbf{Variable} & \textbf{e\textsuperscript{Coef.} (SE)} & \textbf{95\% CI} & \textbf{e\textsuperscript{Coef.}} & \textbf{\% Diff from White} \\
\hline
\% Asian (Compartmental ref.) & 0.463\textsuperscript{***} (0.019) & [0.446, 0.481] & 0.463\textsuperscript{***} & -53.7\% \\
\quad × Baseline Models & 1.635\textsuperscript{***} (0.046) & [1.494, 1.789] & 0.757\textsuperscript{***} & -24.3\% \\
\quad × Deep Learning & 1.549\textsuperscript{***} (0.042) & [1.427, 1.681] & 0.717\textsuperscript{***} & -28.3\% \\
\quad × Ensemble & 1.232\textsuperscript{***} (0.027) & [1.168, 1.300] & 0.571\textsuperscript{***} & -42.9\% \\
\quad × Statistical & 0.934\textsuperscript{*} (0.033) & [0.876, 0.995] & 0.432\textsuperscript{***} & -56.8\% \\
\hline
\% Black (Compartmental ref.) & 1.009\textsuperscript{*} (0.004) & [1.001, 1.017] & 1.009\textsuperscript{*} & +0.9\% \\
\quad × Baseline Models & 1.012 (0.008) & [0.995, 1.028] & 1.021\textsuperscript{*} & +2.1\% \\
\quad × Deep Learning & 0.902\textsuperscript{***} (0.006) & [0.890, 0.913] & 0.910\textsuperscript{***} & -9.0\% \\
\quad × Ensemble & 0.942\textsuperscript{***} (0.005) & [0.933, 0.950] & 0.950\textsuperscript{***} & -5.0\% \\
\quad × Statistical & 0.951\textsuperscript{***} (0.005) & [0.941, 0.960] & 0.959\textsuperscript{***} & -4.1\% \\
\hline
\% Hispanic (Compartmental ref.) & 1.235\textsuperscript{***} (0.004) & [1.226, 1.243] & 1.235\textsuperscript{***} & +23.5\% \\
\quad × Baseline Models & 1.037\textsuperscript{***} (0.008) & [1.021, 1.053] & 1.281\textsuperscript{***} & +28.1\% \\
\quad × Deep Learning & 0.966\textsuperscript{***} (0.006) & [0.955, 0.978] & 1.193\textsuperscript{***} & +19.3\% \\
\quad × Ensemble & 0.958\textsuperscript{***} (0.004) & [0.950, 0.967] & 1.183\textsuperscript{***} & +18.3\% \\
\quad × Statistical & 0.982\textsuperscript{***} (0.005) & [0.972, 0.991] & 1.213\textsuperscript{***} & +21.3\% \\
\hline
\end{tabular}}
{\scriptsize \textbf{Model Statistics:} Pseudo $R^2$ (CS) = 0.461; Log-Likelihood = 5,982,160; N = 1,526,869} 

{\scriptsize \textbf{Notes:} \textsuperscript{***}$p<0.001$, \textsuperscript{**}$p<0.01$, \textsuperscript{*}$p<0.05$. \textit{Dependent Variable:} Square root PBL. \textit{Link Function:} Log. \textit{Regression Family:} Gaussian. The table shows the GLM coefficients and their significance for model GLM-1c. We only discuss race, ethnicity and model type interaction coefficients. For clarity purposes, all other main effects and control variables: Health outcomes, age 65+ and state fixed effects are only shown and discussed in Tables 11, 12 in \nameref{s1appendix}. Model Diagnostics are provided in Fig 5c in \nameref{s1appendix}. For the Coefficient Estimates, \textit{exp(Coef)} represents the multiplicative effect on the outcome, SE: Standard Error and CI: Confidence Interval.
The Relative Effect represents the multiplicative effect on the forecast error (PBL) of a particular race or ethnicity compared to the White population within each model type.
The relative effect is represented by \textit{exp(Coef)} and computed as $e^{\beta_{i} + \delta_{ij}}$ with coefficients from Equation \ref{eq3}.
To evaluate better the relative effect, we also discuss the percentage change in forecast error when compared to White population for each model type (\textit{\% Diff from White}). This change is computed as $(1- e^{\beta_{i} + \delta_{ij}})*100\%$, for a given race/ethnicity and model type, and it represents the percentage increase or decrease in the forecast error (PBL) with respect to the White population (e.g., +23.5\% means that the PBL error for Hispanic counties for Compartmental models is 23.5\% higher PBL when compared to White). All effects should be interpreted as the relative difference compared to White population within each specific model type. The relative coefficient significance is evaluated using \texttt{linearHypothesis} in R (\textit{car} package \cite{fox2018r})
}
\end{table*}

\added{Our analysis of how prediction disparities vary across different minority race/ethnic groups and model types with respect to the White reference group reveals distinct patterns and demonstrates that model architecture choices have significant implications for prediction fairness (see Table \ref{tab:glm-1c}). }

\added{Increases in Hispanic population with respect to the White reference group are associated with large performance disparities across all model types. Baseline models show the largest disparity (28.1\% higher PBL compared to White areas), while Compartmental models show somewhat reduced, though still substantial, disparities (+30.7\%). Ensemble models register the lowest disparity (+18.3\%).}

\added{Increases in Black population are associated with relatively modest variations in PBL across model types when compared to the White reference group. Compartmental and Baseline Models show marginally higher prediction errors (0.9\% \& +2.1\% difference with White areas), while Deep Learning \& Ensemble models demonstrate better relative performance (-9.0\%, -5\%). Statistical models perform better for Black communities (-4.1\%).}

\added{The Asian subgroup, on the other hand, shows improved relative effect with respect to White populations, and these disparities remain relatively consistent across model types, with PBL errors being at least 24.3\% lower than errors for the White group across all model types. }

\added{These findings suggest that model architecture choices significantly impact prediction fairness, with Deep Learning and Ensemble models providing the most balanced performance across racial and ethnic groups, particularly for Hispanic and Black populations }

%\begin{tcolorbox}[title=Summary, colback=white, colframe=black]
%\added{
%The analysis of model type interactions reveals that while model architecture choices can influence prediction fairness, these effects are relatively modest compared to the underlying disparities across racial and ethnic groups. The persistent nature of these disparities across different model types suggests that the challenges in achieving prediction fairness may be more fundamentally related to data collection, underlying social factors, or systematic biases in reporting systems rather than purely algorithmic considerations. The findings highlight the importance of considering fairness implications when selecting model architectures, while also acknowledging that addressing prediction disparities may require attention to broader systemic factors beyond model choice alone.}
%\end{tcolorbox}

\subsubsection*{GLM-1d: Race, Ethnicity and Mobility Data}

\added{Analysis of how prediction disparities vary with mobility data usage also reveal interesting patterns in the relationship between data inputs and prediction fairness (Table \ref{tab:glm-1d}).}

%% Mobility Used X Race
\begin{table*}[h]
\caption{\textbf{GLM-1d}: Race × Mobility Data Inclusion Effects Relative to White }
\label{tab:glm-1d}
\resizebox{\linewidth}{!}{\begin{tabular}{lcc|cc}
\hline
& \multicolumn{2}{c|}{\textbf{Coefficient Estimates}} & \multicolumn{2}{c}{\textbf{Relative Effect}} \\
\textbf{Variable} & \textbf{e\textsuperscript{Coef.} (SE)} & \textbf{95\% CI} & \textbf{e\textsuperscript{Coef.}} & \textbf{\% Diff from White} \\
\hline
\% Asian (No Mobility ref.) & 0.351\textsuperscript{***} (0.027) & [0.332, 0.370] & 0.351\textsuperscript{***} & -64.9\% \\
\quad × Mixed & 1.585\textsuperscript{***} (0.035) & [1.480, 1.697] & 0.556\textsuperscript{***} & -44.4\% \\
\quad × Mobility Used & 1.626\textsuperscript{***} (0.029) & [1.536, 1.722] & 0.571\textsuperscript{***} & -42.9\% \\
\hline
\% Black (No Mobility ref.) & 1.068\textsuperscript{***} (0.005) & [1.058, 1.079] & 1.068\textsuperscript{***} & +6.8\% \\
\quad × Mixed & 0.909\textsuperscript{***} (0.006) & [0.899, 0.919] & 0.971\textsuperscript{***} & -2.9\% \\
\quad × Mobility Used & 0.889\textsuperscript{***} (0.005) & [0.881, 0.897] & 0.949\textsuperscript{***} & -5.1\% \\
\hline
\% Hispanic (No Mobility ref.) & 1.312\textsuperscript{***} (0.005) & [1.301, 1.324] & 1.312\textsuperscript{***} & +31.2\% \\
\quad × Mixed & 0.930\textsuperscript{***} (0.005) & [0.920, 0.940] & 1.220\textsuperscript{***} & +22.0\% \\
\quad × Mobility Used & 0.903\textsuperscript{***} (0.004) & [0.895, 0.911] & 1.185\textsuperscript{***} & +18.5\% \\
\hline
\end{tabular}}
{\scriptsize \textbf{Model Statistics:} Pseudo $R^2$ (CS) = 0.461; Log-Likelihood = 5,982,399; N = 1,526,869} 

{\scriptsize \textbf{Notes:} \textsuperscript{***}$p<0.001$, \textsuperscript{**}$p<0.01$, \textsuperscript{*}$p<0.05$. \textit{Dependent Variable:} Square root  PBL. \textit{Link Function:} Log. \textit{Regression Family:} Gaussian. The table shows the GLM coefficients and their significance for model GLM-1d. We only discuss race, ethnicity and mobility usage interaction coefficients. For clarity purposes, all other main effects and control variables: Health outcomes, age 65+ and state fixed effects are only shown and discussed Tables 13, 14 in  \nameref{s1appendix}. Model Diagnostics are provided in Fig 5d in \nameref{s1appendix}.
For the Coefficient Estimates, \textit{exp(Coef)} represents the multiplicative effect on the outcome,
SE: Standard Error and CI: Confidence Interval.
The Relative Effect represents the multiplicative effect on the forecast error (PBL) of a particular race or ethnicity compared to the White population within each mobility usage category.
The relative effect is represented by \textit{exp(Coef)} and computed as $e^{\beta_{i} + \delta_{ij}}$ with coefficients from Equation \ref{eq3}.
To evaluate better the relative effect, we also discuss the percentage change in forecast error when compared to White population for each mobility usage category (\textit{\% Diff from White}).
This change is computed as $(1- e^{\beta_{i} + \delta_{ij}})*100\%$, for a given race/ethnicity and mobility usage category, and it represents the percentage increase or decrease in the forecast error (PBL) with respect to the White population (e.g., +31.2\% means that the PBL error for Hispanic counties for the given mobility usage category is 31.2\% higher PBL when compared to White). All effects should be interpreted as the relative difference compared to White population within each specific mobility usage category. The relative coefficient significance is evaluated using \texttt{linearHypothesis} in R (\textit{car} package \cite{fox2018r}) }
\end{table*}

%\added{For Hispanic populations, models without mobility shows substantially higher prediction errors ($\exp(\beta)$ = 1.312, $p < 0.001$). While the other two categories showed some moderation of these disparities (Not Applicable: $\exp(\beta)$ = 0.930; Mobility Used: $\exp(\beta)$ = 0.903; both $p < 0.001$). 
\added{The relative effect analysis for Hispanic population indicates significant disparities in forecast performance when compared to the White reference group across mobility data uses in prediction models.
%looking into the interaction between race and the use of mobility data in the prediction models. 
For forecast models that use mobility data, increases in Hispanic population with respect to the White reference group 
are associated with a PBL 18.5\% higher than that of the White population.
That number increases to 22\% for mixed models (ensembles of CDC ForecastHub models trained with and without mobility data).
On the other hand, when mobility data is not used, PBL errors are 31.2\% higher for each 1\% increase in Hispanic population with respect to the White reference group, 
%and when mobility data is ''not applicable'' 
%while models using mobility data show a slightly reduced but still substantial disparity of 18.5\%. 
suggesting that mobility data usage can improve prediction fairness for Hispanic communities. However, considerable disparities remain (higher PBL errors) when compared to the White population.} 

\added{For Black population, and when compared to the White reference group, not using mobility data in the forecast models was associated with PBL errors being 6.8\% higher. However, using mobility data reversed that trend, with PBL errors being 5.1\% lower when the Black population increases 1\% with respect to the White group. }

\added{Prediction performance for Asian populations when considering the use of mobility data was significantly better than the performance for White counties, independently of whether mobility data was used or not in the models, with PBL errors being between 42.9\% and 64.9\% significantly lower than errors for the White reference group. Interestingly, we also observe that use of mobility data helps reduce the disparity in performance errors (42.9\% and 44.4\% vs. 64.9\%). }
%The relative effect of Similarly to other minority races, we also observe that imilar patterns were seen while observing the relative effects as well, with mobility data helping to reduce the disparity. }

\added{These results reveal significant prediction performance differences between minority racial and ethnic groups with respect to White across mobility data use settings; while suggesting that the incorporation of mobility data might help reduce prediction disparities for minority racial groups.}

%predominantly Black communities, models without mobility data register a slight positive bias in predictions ($\exp(\beta)$ = 1.068, $p < 0.001$). However, the bias is overturned for both ``Not Applicable" and ``Mobility Used" categories. Models where mobility data usage was not applicable show 2.9\% lower PBL compared to White areas, while models actively using mobility data demonstrate a slightly larger reduction of 5.1\%. This suggests that the incorporation of mobility data might help reduce prediction disparities for Black communities.}

%\todo{Add a comprehensive summary}
%\vanessa{When writing the summary, I would just focus on the results discussed here, and add some text in the Discussion hypothesizing WHY these results might be happening related to data, struc diffs. add a final sentence saying "In the Discussion section, we will discuss potential reasons behind these findings".
%}

\begin{tcolorbox}[title=\textbf{Summary}: Race X Model-Data Characteristics, colback=gray!10, colframe=gray!50]
\begin{itemize}
    \setlength\itemsep{0.0001em}
    \renewcommand{\labelitemi}{$\Rightarrow$}    
    \item Hispanic communities consistently experience higher prediction errors compared to White areas across most configurations. 
    \item Disparities decrease with longer forecast horizons.
    \item Initial phases of the pandemic saw widest disparity for both Hispanic and Black subgroups
    \item Asian communities maintained lower error rates overall \& showed substantial phase-dependent variations (-88.9\% to +387.1\%)
    \item Deep Learning and Ensemble models demonstrated most balanced performance across all racial and ethnic groups
    \item Mobility data usage generally helps reduce prediction disparities
\end{itemize}
\end{tcolorbox}

%\begin{tcolorbox}[title=Summary, colback=white, colframe=black]
%\added{
%The analysis of mobility data usage interactions reveals that while the incorporation of mobility data can influence prediction fairness, its effects are relatively modest and vary across racial and ethnic groups. The findings suggest that while mobility data might help reduce some disparities, particularly for Black communities, it does not fundamentally resolve the underlying patterns of PBL. This underscores the importance of considering multiple approaches to improving prediction fairness, as data input choices alone may not be sufficient to address systematic disparities in prediction accuracy.}
%\end{tcolorbox}

\subsection*{Fairness of COVID-19 case predictions across urbanicity and model-data characteristics}
%% Urbanicity X Lookahead
\begin{table*}[h]
\caption{\centering \textbf{GLM-2a}: Urbanicity × Lookahead Effects Relative to LM Reference Group}
\label{tab:glm-2a}
\resizebox{\linewidth}{!}{\begin{tabular}{lcc|cc}
\hline
& \multicolumn{2}{c|}{\textbf{Coefficient Estimates}} & \multicolumn{2}{c}{\textbf{Relative Effect}} \\
\textbf{Variable} & \textbf{e\textsuperscript{Coef.} (SE)} & \textbf{95\% CI} & \textbf{e\textsuperscript{Coef.}} & \textbf{\% Diff from LM} \\
\hline
\textbf{Micropolitan (MC)} (7-day ref.) & 1.133\textsuperscript{***} (0.002) & [1.128, 1.137] & 1.133\textsuperscript{***} & +13.3\% \\
\quad × 14-day ahead & 0.963\textsuperscript{***} (0.003) & [0.958, 0.968] & 1.091\textsuperscript{***} & +9.1\% \\
\quad × 21-day ahead & 0.926\textsuperscript{***} (0.003) & [0.921, 0.930] & 1.049\textsuperscript{***} & +4.9\% \\
\quad × 28-day ahead & 0.900\textsuperscript{***} (0.002) & [0.896, 0.905] & 1.020\textsuperscript{***} & +2.0\% \\
\hline
\textbf{Small/Medium Metro (SMM)} (7-day ref.)& 1.048\textsuperscript{***} (0.002) & [1.044, 1.053] & 1.048\textsuperscript{***} & +4.8\% \\
\quad × 14-day ahead & 0.984\textsuperscript{***} (0.003) & [0.979, 0.990] & 1.031\textsuperscript{***} & +3.1\% \\
\quad × 21-day ahead & 0.975\textsuperscript{***} (0.003) & [0.970, 0.981] & 1.022\textsuperscript{***} & +2.2\% \\
\quad × 28-day ahead & 0.969\textsuperscript{***} (0.003) & [0.964, 0.975] & 1.015\textsuperscript{***} & +1.5\% \\
\hline
\end{tabular}}
{\scriptsize \textbf{Model Statistics:} Pseudo $R^2$ (CS) = 0.464; Log-Likelihood = 5,985,040; N = 1,526,869} 

{\scriptsize \textbf{Notes:} \textsuperscript{***}$p<0.001$, \textsuperscript{**}$p<0.01$, \textsuperscript{*}$p<0.05$. \textit{Dependent Variable:} Square root PBL. \textit{Link Function:} Log. \textit{Regression Family:} Gaussian. The table shows the GLM coefficients and their significance for model GLM-2a. We only discuss urbanicity and lookahead interaction coefficients. For clarity purposes, all other main effects and control variables: Health outcomes, age 65+ and state fixed effects are only shown and discussed in Tables 15, 16 in  \nameref{s1appendix}. Model Diagnostics are provided in Fig 6a in \nameref{s1appendix}. For the Coefficient Estimates, \textit{exp(Coef)} represents the multiplicative effect on the outcome, SE: Standard Error and CI: Confidence Interval.
The Relative Effect represents the multiplicative effect on the forecast error (PBL) of a particular urbanicity level compared to the Large Metropolitan (LM) areas within each lookahead.
The relative effect is represented by \textit{exp(Coef)} and computed as $e^{\beta_{i} + \delta_{ij}}$ with coefficients from Equation \ref{eq:eq4}.
To evaluate better the relative effect, we also discuss the percentage change in forecast error when compared to LM areas for each lookahead variable (\textit{\% Diff from LM}).
This change is computed as $(1- e^{\beta_{i} + \delta_{ij}})*100\%$
for a given urbanicity level and lookahead value, and it represents the percentage increase or decrease in the forecast error (PBL) with respect to LM areas. All effects should be interpreted as the relative difference compared to LM areas within each specific lookahead value. The relative coefficient significance is evaluated using \texttt{linearHypothesis} in R (\textit{car} package \cite{fox2018r}) }
\vspace{0.2cm}

\end{table*}

%% Urbanicity X Phase
\begin{table*}[h]
\caption{\textbf{GLM-2b}: Urbanicity × Phase Effects Relative to LM Areas}
\label{tab:glm-2b}
\resizebox{\linewidth}{!}{\begin{tabular}{lcc|cc}
\hline
& \multicolumn{2}{c|}{\textbf{Coefficient Estimates}} & \multicolumn{2}{c}{\textbf{Relative Effect}} \\
\textbf{Variable} & \textbf{e\textsuperscript{Coef.} (SE)} & \textbf{95\% CI} & \textbf{e\textsuperscript{Coef.}} & \textbf{\% Diff from LM} \\
\hline
\textbf{Micropolitan (MC)} (Phase 0 ref.) & 1.116\textsuperscript{***} (0.003) & [1.110, 1.122] & 1.116\textsuperscript{***} & +11.6\% \\
\quad × Phase 1 & 0.997 (0.003) & [0.991, 1.003] & 1.113\textsuperscript{***} & +11.3\% \\
\quad × Phase 2 & 0.858\textsuperscript{***} (0.004) & [0.852, 0.864] & 0.958\textsuperscript{***} & -4.2\% \\
\quad × Phase 3 & 1.007\textsuperscript{*} (0.003) & [1.001, 1.014] & 1.124\textsuperscript{***} & +12.4\% \\
\quad × Phase 4 & 0.915\textsuperscript{***} (0.003) & [0.909, 0.921] & 1.021\textsuperscript{***} & +2.1\% \\
\quad × Phase 5 & 0.820\textsuperscript{***} (0.004) & [0.814, 0.827] & 0.915\textsuperscript{***} & -8.5\% \\
\quad × Phase 6 & 1.004 (0.004) & [0.996, 1.013] & 1.120\textsuperscript{***} & +12.0\% \\
\hline
\textbf{Small/Medium Metro (SMM)} (Phase 0 ref.) & 1.091\textsuperscript{***} (0.003) & [1.085, 1.098] & 1.091\textsuperscript{***} & +9.1\% \\
\quad × Phase 1 & 0.960\textsuperscript{***} (0.004) & [0.953, 0.966] & 1.047\textsuperscript{***} & +4.7\% \\
\quad × Phase 2 & 0.876\textsuperscript{***} (0.004) & [0.869, 0.883] & 0.956\textsuperscript{***} & -4.4\% \\
\quad × Phase 3 & 0.998 (0.004) & [0.991, 1.006] & 1.089\textsuperscript{***} & +8.9\% \\
\quad × Phase 4 & 0.898\textsuperscript{***} (0.004) & [0.891, 0.904] & 0.980\textsuperscript{***} & -2.0\% \\
\quad × Phase 5 & 0.862\textsuperscript{***} (0.005) & [0.854, 0.870] & 0.940\textsuperscript{***} & -6.0\% \\
\quad × Phase 6 & 0.971\textsuperscript{***} (0.005) & [0.963, 0.980] & 1.059\textsuperscript{***} & +5.9\% \\
\hline
\end{tabular}}
{\scriptsize \textbf{Model Statistics:} Pseudo $R^2$ (CS) = 0.466; Log-Likelihood = 5,987,062; N = 1,526,869} \\
{\scriptsize \textbf{Notes:} \textsuperscript{***}$p<0.001$, \textsuperscript{**}$p<0.01$, \textsuperscript{*}$p<0.05$. \textit{Dependent Variable:} Square root PBL. \textit{Link Function:} Log. \textit{Regression Family:} Gaussian. The table shows the GLM coefficients and their significance for model GLM-2b. We only discuss urbanicity and phase interaction coefficients. For clarity purposes, all other main effects and control variables: Health outcomes, age 65+ and state fixed effects are only shown and discussed in Tables 17, 18 in  \nameref{s1appendix}. Model Diagnostics are provided in Fig 6b in \nameref{s1appendix}.
For the Coefficient Estimates, \textit{exp(Coef)} represents the multiplicative effect on the outcome,
SE: Standard Error and CI: Confidence Interval.
The Relative Effect represents the multiplicative effect on the forecast error (PBL) of a particular urbanicity level compared to the Large Metropolitan (LM) areas within each phase.
The relative effect is represented by \textit{exp(Coef)} and computed as $e^{\beta_{i} + \delta_{ij}}$ with coefficients from Equation \ref{eq:eq4}.
To evaluate better the relative effect, we also discuss the percentage change in forecast error when compared to LM areas for each phase (\textit{\% Diff from LM}).
This change is computed as $(1- e^{\beta_{i} + \delta_{ij}})*100\%$
for a given urbanicity level and phase, and it represents the percentage increase or decrease in the forecast error (PBL) with respect to LM areas. All effects should be interpreted as the relative difference compared to LM areas within each specific phase. The relative coefficient significance is evaluated using \texttt{linearHypothesis} in R (\textit{car} package \cite{fox2018r}) }
\end{table*}

\subsubsection*{GLM-2a: Urbanicity and Forecast Lookahead}

\added{Table \ref{tab:glm-2a} exhibits significant disparities in COVID-19 case prediction errors between urbanization levels across different forecast horizons. Our findings demonstrate that both MC and SMM areas consistently experience higher prediction errors when compared to LM areas (reference group) across lookaheads; and that the magnitude of these disparities decreases as the prediction horizon extends.}

\added{MC areas show the most pronounced disparity, with baseline (7-day) prediction errors 13.3\% higher than LM areas. This disparity, while persistent, diminishes at longer prediction horizons. For 14-day forecasts, the relative effect shows that MC areas still experience 9.1\% higher errors than LM areas. This gap continues to narrow, decreasing to 4.9\% for 21-day forecasts and further reducing to 2.0\% for 28-day predictions. }

\added{SMM areas exhibit a similar pattern but with smaller magnitudes of disparity compared to MC. These areas show prediction errors 4.8\% higher than LM areas for 7-day forecasts. Again the pattern of the disparity decreasing is also observed for SMMs, with relative effects showing 3.1\% higher errors for 14-day forecasts, 2.2\% for 21-day forecasts, and 1.5\% for 28-day forecasts compared to LM areas. These differences, while smaller, remain statistically significant across all forecast horizons $(p < 0.001)$.}
%\begin{tcolorbox}[title=Summary, colback=white, colframe=black]
%\added{
%The analysis of urbanicity and forecast horizon interactions reveals that while urban-rural disparities in prediction accuracy persist across all forecast horizons, these disparities tend to decrease with longer-term predictions. The convergence pattern is particularly pronounced for micropolitan areas, where the initial larger disparity reduces substantially over longer forecast horizons. This suggests that while short-term predictions show marked urban-rural differences, longer-term forecasts achieve more consistent accuracy across different settlement patterns, potentially due to the averaging out of short-term fluctuations in case reporting and detection.}
%\end{tcolorbox}

\subsubsection*{GLM-2b: Urbanicity and COVID-19 Phase}

\added{
Both MC and SMM areas show significantly different prediction performance compared to LM areas, with these disparities varying substantially across different pandemic phases (See Table \ref{tab:glm-2b}).}

\added{
MC areas demonstrate the highest baseline disparity (Phase 0), with prediction errors 11.6\% higher than LM areas. The magnitude of this disparity fluctuates notably across different pandemic phases. During Phase 3, MC areas experienced their largest disparity, with errors 12.4\% higher than LM areas. Conversely, Phases 2 and 5 show a reversal of this pattern, with MC areas actually performing better than LM areas, showing 4.2\% and 8.5\% lower errors respectively $(p < 0.001)$. Phases 1 and 6 maintained similar disparities to the baseline (11.3\% and 12.0\% higher errors)}

\added{
SMM areas show a similar but less pronounced pattern of disparities. The baseline prediction errors for SMM areas are 9.1\% higher than LM areas. Like MC areas, SMM areas show varying performance across phases, but with generally smaller magnitudes of disparity. The pattern of better performance in Phases 2 and 5 is repeated, with SMM areas showing 4.4\% and 6.0\% lower errors than LM areas respectively ($p < 0.001$). The highest disparities for SMM areas occur in Phase 3 (8.9\% higher errors) and Phase 6 (5.9\% higher errors).}

%\added{Similary to the racial and ethnic analysis, we observe the highest percentage error differences with the reference group (LM) during phase 3.}%, which is also another COVID-19 peak-case phase (see Figure \ref{fig:2b}). }
%\vanessa{Saad, didn't we have some discussion around how this could be due to data collection issues or to peak cases in certain phases? I think we should add this in the discussion, and in fact, we should add these for all model-tdata characteristics trying to provide hypothesis about why this might be happening.}

%\subsubsection*{GLM 2b: Urbanicity X Phase}

\subsubsection*{GLM-2c Urbanicity and Model Type}
%% Urbanicity X Model Type
\begin{table*}[h]
\caption{\textbf{GLM-2c}: Urbanicity × Model Type Effects Relative to LM Areas}
\label{tab:glm-2c}
\resizebox{\linewidth}{!}{\begin{tabular}{lcc|cc}
\hline
& \multicolumn{2}{c|}{\textbf{Coefficient Estimates}} & \multicolumn{2}{c}{\textbf{Relative Effect}} \\
\textbf{Variable} & \textbf{e\textsuperscript{Coef.} (SE)} & \textbf{95\% CI} & \textbf{e\textsuperscript{Coef.}} & \textbf{\% Diff from LM} \\
\hline
\textbf{Micropolitan (MC)} (Compartmental ref.) & 1.066\textsuperscript{***} (0.001) & [1.063, 1.069] & 1.066\textsuperscript{***} & +6.6\% \\
\quad × Baseline Models & 0.984\textsuperscript{***} (0.004) & [0.977, 0.991] & 1.049\textsuperscript{***} & +4.9\% \\
\quad × Deep Learning & 0.979\textsuperscript{***} (0.003) & [0.974, 0.985] & 1.044\textsuperscript{***} & +4.4\% \\
\quad × Ensemble & 0.988\textsuperscript{***} (0.002) & [0.984, 0.992] & 1.053\textsuperscript{***} & +5.3\% \\
\quad × Statistical & 1.029\textsuperscript{***} (0.002) & [1.024, 1.034] & 1.097\textsuperscript{***} & +9.7\% \\
\hline
\textbf{Small/Medium Metro (SMM)} (Compartmental ref.) & 1.028\textsuperscript{***} (0.001) & [1.025, 1.031] & 1.028\textsuperscript{***} & +2.8\% \\
\quad × Baseline Models & 1.000 (0.004) & [0.992, 1.009] & 1.028\textsuperscript{***} & +2.8\% \\
\quad × Deep Learning & 0.983\textsuperscript{***} (0.003) & [0.976, 0.989] & 1.011\textsuperscript{***} & +1.1\% \\
\quad × Ensemble & 0.990\textsuperscript{***} (0.002) & [0.986, 0.995] & 1.018\textsuperscript{***} & +1.8\% \\
\quad × Statistical & 1.019\textsuperscript{***} (0.003) & [1.014, 1.025] & 1.048\textsuperscript{***} & +4.8\% \\
\hline
\end{tabular}}
{\scriptsize \textbf{Model Statistics:} Pseudo $R^2$ (CS) = 0.462; Log-Likelihood = 5,983,479; N = 1,526,869} \\
{\scriptsize \textbf{Notes:} \textsuperscript{}$p<0.001$, \textsuperscript{}$p<0.01$, \textsuperscript{}$p<0.05$. \textit{Dependent Variable:} Square root PBL. \textit{Link Function:} Log. \textit{Regression Family:} Gaussian. The table shows the GLM coefficients and their significance for model GLM-2c. We only discuss urbanicity and model type interaction coefficients. For clarity purposes, all other control variables: Health outcomes, age 65+ and state fixed effects are only shown and discussed in Tables 19, 20 in  \nameref{s1appendix}. Model Diagnostics are provided in Fig 6c in \nameref{s1appendix}.
For the Coefficient Estimates, \textit{exp(Coef)} represents the multiplicative effect on the outcome,
SE: Standard Error and CI: Confidence Interval.
The Relative Effect represents the multiplicative effect on the forecast error (PBL) of a particular urbanicity level compared to the Large Metropolitan (LM) areas within each model type.
The relative effect is represented by \textit{exp(Coef)} and computed as $e^{\beta_{i} + \delta_{ij}}$ with coefficients from Equation \ref{eq:eq4}.
To evaluate better the relative effect, we also discuss the percentage change in forecast error when compared to LM areas for each model type (\textit{\% Diff from LM}).
This change is computed as $(1- e^{\beta_{i} + \delta_{ij}})*100\%$
for a given urbanicity level and model type, and it represents the percentage increase or decrease in the forecast error (PBL) with respect to LM areas. All effects should be interpreted as the relative difference compared to LM areas within each specific model type. The relative coefficient significance is evaluated using \texttt{linearHypothesis} in R (\textit{car} package \cite{fox2018r}) }
\vspace{0.2cm}

\end{table*}

\added{Table \ref{tab:glm-2c} shows the interaction effect analysis for urbanicity levels and types of forecast models. Overall, we observe that MC and SMM suffer from higher PBL errors than LM areas across all types of forecasting models. }

\added{The highest performance disparities across model types are observed between MC and LM counties. Specifically, statistical models are the ones with the most pronounced differences, with MC areas experiencing 9.7\% higher errors than LM areas;
followed by Compartmental models (reference group) at 6.6\%, 
Ensemble models at 5.3\%, baseline models at 4.9\% and deep learning models at 4.4\%. All these differences remain statistically significant ($p < 0.001$).}

%, as well as the interaction and selected main effects. We observe very similar patterns, with MC areas showing a baseline disparity of 6.6\% higher prediction errors compared to LM areas ($exp(\beta) = 1.066, p < 0.001$). This disparity is most pronounced when statistical models are used, with MC areas experiencing 9.7\% higher errors than LM areas. Ensemble models show the second-highest disparity at 5.3\%, followed by baseline models at 4.9\% and deep learning models at 4.4\%. All these differences remain statistically significant ($p < 0.001$).}

\added{As seen before for prediction lookaheads and some phases, SMM areas demonstrate a similar pattern but with smaller magnitudes of disparity. Statistical models show the largest disparity between SMM and LM counties, with PBL errors being 4.8\% higher in SMM areas. Compartmental (reference group) and baseline model prediction errors for SMM counties are 2.8\% higher than LM areas. The disparities are notably smaller for ensemble and deep learning models. Ensemble models have PBL errors 1.8\% higher in SMM counties when compared to LM counties, while deep learning models forecasting cases in SMM counties have PBL errors 1.1\% higher than LMs. }

\added{These findings suggest that deep learning models may be most effective at minimizing urbanization-related disparities in prediction accuracy. Conversely, statistical models appear to amplify these disparities across both MC and SMM areas. The consistent pattern of higher disparities in MC areas compared to SMM areas, regardless of model type, indicates that predictive challenges in less urbanized areas persist across modeling approaches, though their magnitude can be influenced by model selection.}

%\vanessa{Saad, as I mentioned for phases and lookaheads, I think it might make more sense to move these findings to the Discussion, or maybe in the summary box at the end of each section?}

\subsubsection*{\textbf{GLM-2d}: Urbanicity and Mobility Data}

%% Urbanicity X Mobility Used
\begin{table*}[h]
\caption{\textbf{GLM-2d}: Urbanicity × Mobility Usage Effects Relative to LM}
\label{tab:glm-2d}
\resizebox{\linewidth}{!}{\begin{tabular}{lcc|cc}
\hline
& \multicolumn{2}{c|}{\textbf{Coefficient Estimates}} & \multicolumn{2}{c}{\textbf{Relative Effect}} \\
\textbf{Variable} & \textbf{e\textsuperscript{Coef.} (SE)} & \textbf{95\% CI} & \textbf{e\textsuperscript{Coef.}} & \textbf{\% Diff from Rural} \\
\hline
\textbf{Micropolitan (MC)} (No Mobility ref.) & 1.088\textsuperscript{***} (0.002) & [1.084, 1.092] & 1.088\textsuperscript{***} & +8.8\% \\
\quad × Mixed & 0.979\textsuperscript{***} (0.003) & [0.974, 0.984] & 1.065\textsuperscript{***} & +6.5\% \\
\quad × Mobility Used & 0.972\textsuperscript{***} (0.002) & [0.968, 0.976] & 1.058\textsuperscript{***} & +5.8\% \\
\hline
\textbf{Small/Medium Metro(SMM)} (No Mobility ref.) & 1.056\textsuperscript{***} (0.002) & [1.052, 1.061] & 1.056\textsuperscript{***} & +5.6\% \\
\quad × Mixed & 0.976\textsuperscript{***} (0.003) & [0.970, 0.981] & 1.031\textsuperscript{***} & +3.1\% \\
\quad × Mobility Used & 0.963\textsuperscript{***} (0.002) & [0.958, 0.967] & 1.017\textsuperscript{***} & +1.7\% \\
\hline
\end{tabular}}
{\scriptsize \textbf{Model Statistics:} Pseudo $R^2$ (CS) = 0.462; Log-Likelihood = 5,983,425; N = 1,526,869} \\
{\scriptsize \textbf{Notes:} \textsuperscript{***}$p<0.001$, \textsuperscript{**}$p<0.01$, \textsuperscript{*}$p<0.05$. \textit{Dependent Variable:} Square root PBL. \textit{Link Function:} Log. \textit{Regression Family:} Gaussian. The table shows the GLM coefficients and their significance for model GLM-2d. We only discuss urbanicity and mobility usage interaction coefficients. For clarity purposes, all other control variables: Health outcomes, age 65+ and state fixed effects are only shown and discussed in Tables 21, 22 in  \nameref{s1appendix}. Model Diagnostics are provided in Fig 6d in \nameref{s1appendix}.
For the Coefficient Estimates, \textit{exp(Coef)} represents the multiplicative effect on the outcome,
SE: Standard Error and CI: Confidence Interval. The Relative Effect represents the multiplicative effect on the forecast error (PBL) of a particular urbanicity level compared to the Large Metropolitan (LM) areas within each mobility usage category. The relative effect is represented by \textit{exp(Coef)} and computed as $e^{\beta_{i} + \delta_{ij}}$ with coefficients from Equation \ref{eq:eq4}. To evaluate better the relative effect, we also discuss the percentage change in forecast error when compared to LM areas for each mobility usage category (\textit{\% Diff from LM}).
This change is computed as $(1- e^{\beta_{i} + \delta_{ij}})*100\%$
for a given urbanicity level and mobility usage category, and it represents the percentage increase or decrease in the forecast error (PBL) with respect to LM areas. All effects should be interpreted as the relative difference compared to LM areas within each specific mobility usage category. The relative coefficient significance is evaluated using \texttt{linearHypothesis} in R (\textit{car} package \cite{fox2018r}) }
\end{table*}

\added{Table \ref{tab:glm-2d} represents the interaction analysis between urbanization levels and the use of mobility data in forecasting models. The relative effects analyses show a similar pattern to the other model-data characteristics we
have discussed \textit{i.e.,} when compared against LM counties, MC and SMM counties are associated with significantly higher prediction errors across mobility data use approaches (no mobility data, mobility data or mixed model). }

\added{MC areas exhibit the largest disparity, with prediction errors 8.8\% higher than LM areas when no mobility data is used to train the COVID-19 case forecasting models. This disparity is modestly reduced when mobility data is used, with MC areas showing 5.8\% higher errors compared to LM areas, and 6.5\% higher errors for mixed models (ensemble of models trained with and without mobility data). Both reductions in disparity are statistically significant $(p < 0.001)$.}

\added{On the other hand, SMM areas demonstrate a similar pattern but with smaller magnitudes of disparity. Not using mobility causes prediction errors for SMM areas to be 5.6\% higher than LM areas. The incorporation of mobility data appears to be more effective in reducing disparities for SMM areas as well, with models using mobility data showing only 1.7\% higher errors compared to LM areas. Mixed models, based on CDC ForecastHub ensembles of both models trained with and without mobility data show an intermediate improvement, with 3.1\% higher errors compared to LM areas.}

\begin{tcolorbox}[title=\textbf{Summary}: Urbanicity X Model-Data Characteristics, colback=gray!10, colframe=gray!50]

\begin{itemize}
    \setlength\itemsep{0.0001em}
    \renewcommand{\labelitemi}{$\Rightarrow$}    
    \item SMM and MC are consistently associated to higher prediction errors compared to LM across most configurations. 
    \item Disparities decrease with longer forecast horizons.
    \item Phases 2 and 5 saw a reversal in the pattern with SMM and MC performing better than LM counties.
    \item Deep learning models demonstrated most balanced performance across urbanization levels.
    \item Mobility data usage generally helps reduce prediction disparities
\end{itemize}

%Our analysis reveals consistent disparities in COVID-19 forecast accuracy across urbanization levels, with both MC and SMM areas experiencing higher prediction errors compared to LM areas. For forecast horizons, MC areas show the largest disparities (13.3\% higher errors for 7-day forecasts), decreasing to 2.0\% at 28 days, while SMM areas follow a similar but less pronounced pattern (4.8\% to 1.5\%). Across pandemic phases, disparities fluctuate significantly, with Phase 3 showing the highest disparities for both MC (12.4\%) and SMM (8.9\%) areas. Model architecture analysis reveals that Deep Learning models minimize urbanization-related disparities (4.4\% for MC, 1.1\% for SMM), while Statistical models show the largest disparities (9.7\% for MC, 4.8\% for SMM). The use of mobility data generally reduces disparities, particularly for SMM areas (from 5.6\% to 1.7\%).

\end{tcolorbox}

%\todo{consolidated summary}
%\vanessa{When writing the summary,focus on results discussed here, and add a final sentence saying "In the Discussion section, we will discuss potential reasons behind these findings".}

\section*{Dashboard}

%This paper presents a thorough analysis of the Forecast Hub's COVID-19 county case prediction models' performance across race, ethnicity and urbanization level.

%and we delve into fairness changes when 

%. We carry out statistical analyses identifying differences in performance 
%across groups and types of models, datasets, lookaheads and phases. Our research shows statistically significant differences in predictive errors with some minority racial and ethnic groups as well as rural areas associated with significantly higher errors. 

This paper has revealed significant disparities in COVID-19 case prediction accuracy across race, ethnicity and urbanization level. 
%different racial, ethnic, and urbanization groups. 
%These disparities are highlighted by higher prediction errors for certain groups and urbanization levels. 
\added{The regression analyses we have presented, evaluate the relationship between COVID-19 case prediction errors and  racial and ethnic groups, urbanization level, model type, the use of mobility data, lookahead, and phase. 
Our findings are based on global trends across all Forecast Hub models, and provide general recommendations for researchers working in COVID-19 prediction models and for decision makers using case predictions to inform pandemic policies. For example, we have shown that 
mobility data helps reduce the forecast error disparities between racial groups and urbanization levels. 
This fact can be used modelers and decision makers to support the use of mobility data to enhance forecast models. }

%revealed that deep learning models are associated with higher errors for some minority groups and for rural areas. This finding can inform modelers to deepen into the reasons why deep learning models are failing more for certain racial groups, and can also guide decision makers to take predictions from deep learning
%models with a grain of salt when making decisions for rural areas. }

\begin{figure}[!ht]
    \includegraphics[width=\textwidth]{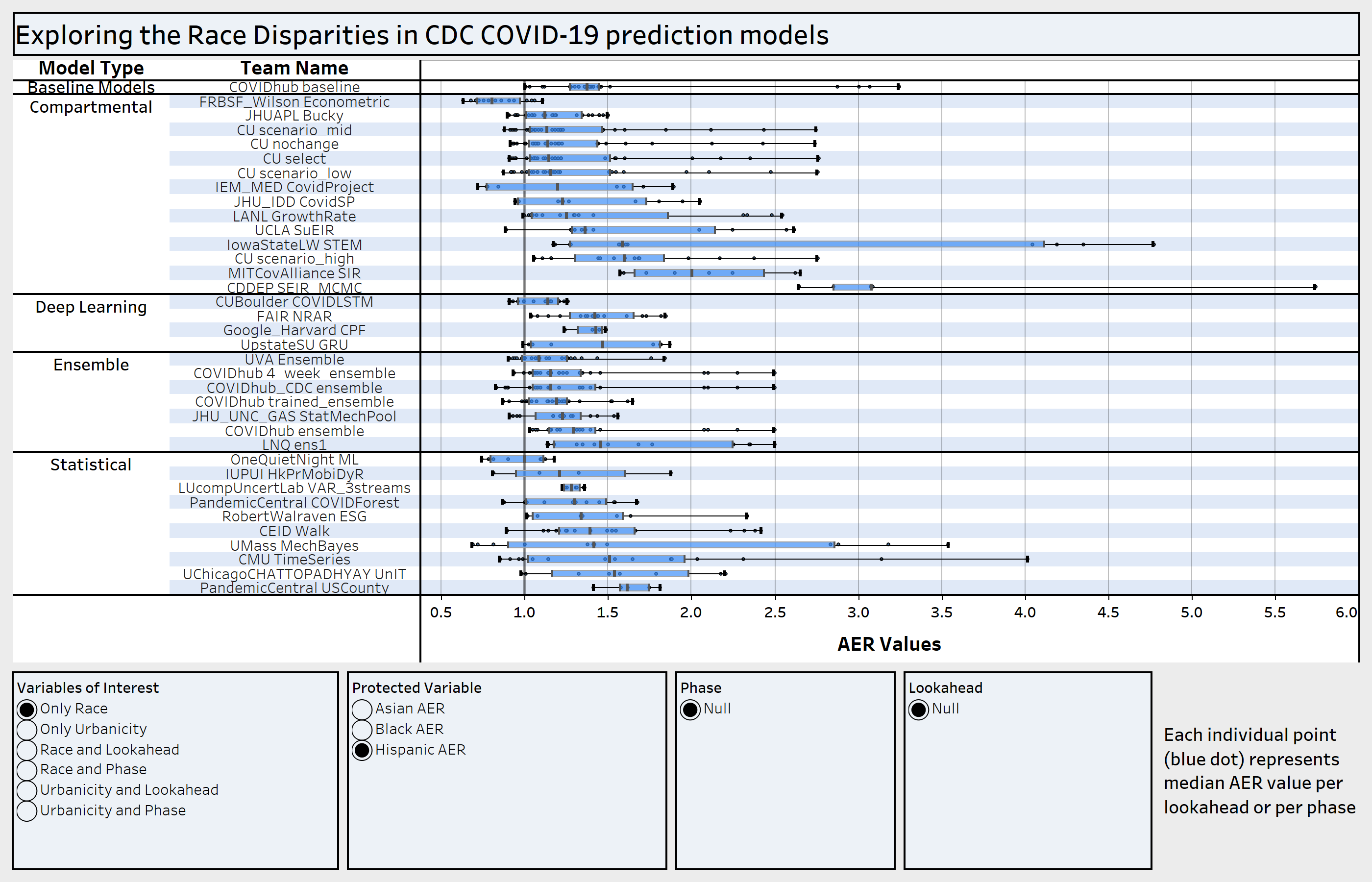}
    \caption{ \added{\textbf{Forecast Hub Fairness Dashboard} showing the Average Error Ratio (AER) distribution across different COVID-19 prediction models, organized by model type. Within each model type, teams are sorted in ascending order based on their median AER values. Since the user has selected ``Only Race" as the variable of interest (see bottom left box) and ``Hispanic" as the protected variable (see bottom center box), the AER values compare prediction errors between Hispanic and White counties, where values above 1.0 indicate higher prediction errors for Hispanic counties. Box plots show the distribution of AER values across all predictions, with the center line representing the median, boxes showing the interquartile range, and whiskers extending to the minimum and maximum values.}
    }
    \label{fig:dashboard}
\end{figure}

Nevertheless, it is important to acknowledge that researchers and decision makers might also want to assess the specific performance of each COVID-19 county case prediction model individually, exploring PBL error differences between racial and ethnic groups or urbanization levels for a given model, their statistical significance, or whether these differences persist when considering specific lookaheads or phases. 
%and in relation to these groups, and examine the significance and magnitude of the differences in prediction accuracy between protected and unprotected groups.
To enable individual model evaluation, we have created an interactive dashboard (see Fig ~\ref{fig:dashboard}), that will be made publicly available upon the publication of this paper. 
The dashboard displays a model's performance error (PBL) for a given protected attribute - race and ethnicity or urbanization level - that can be selected by the user from the user interface.

To allow for meaningful explorations, the individual model errors are displayed using the Accuracy Equality Ratio (AER)~\cite{castelnovo2022clarification}, which measures the difference in error distributions between protected and unprotected groups for a given protected attribute. The AER is computed as a quotient between the model's performance error (PBL) for a given protected group $g$ across all counties and the model's performance error for the unprotected group across all counties:
$AER_g = \frac{ PBL(protected\_ group\_g)}{PBL(unprotected\_ group)}$
where $PBL(protected\_ group\_g)$ and $PBL(unprotected\_ group)$ are the pinball ball loss metric for protected and unprotected groups respectively.
For the race and ethnicity protected attribute we define 
%the following protected groups with respect to White: 
three protected groups with respect to the White unprotected group: Asian, Black and Hispanic, and associate the plurality race to each county, i.e., the race or ethnicity that makes up the largest percentage for that county.  
%Black, Hispanic and Non-White for the majority approach; and Asian, Black and Hispanic for the plurality approach. 
For the urbanization code, we use the protected groups described in the paper: 
%with respect to Large Metropolitan Areas: 
Micropolitan (MC) and Small and Medium Metro Areas (SMM) with Large Metropolitan Areas being the unprotected group. 
%s well as Small and Medium Metro Areas. 
%\vanessa{Saad, in the two sentences above, I have modified this a bit because here we still need the majority assignment, right?}
Similar error distributions between the protected and the unprotected groups will produce AER values close to one. AER values larger than one point to higher errors for the protected group, and AER values smaller than one point to higher errors for White or large metropolitan counties (baseline groups). 
%For the race and ethnicity attribute, we consider as protected classes the minority groups in our study, namely Asian, Black and Hispanic, with White being the unprotected class; while for the urbanization levels, the protected classes are the micropolitan areas as well as the 
%As the Figure shows, for each individual model, researchers can explore the fairness of the model across racial and ethnic groups, as well as across urban-rural codes, with fairness measured as differences in error rates across the protected groups for a given attribute. To be able to compare the performance across racial and ethnic groups, or across urban-rural codes, we use 
%the Accuracy Equality Ratio (AER)~\cite{castelnovo2022clarification}, which measures the differences in error distributions for two protected groups. 

%\vanessa{Saad, aren't the asterisks in the model card repetitive of the "statistical significance = true" field? if so please remove. Also, are upper and lower difference the upper and lower values for the AER or for the PBL error? The plot says AER values at the bottom but I think you mean error? I don't understand that. Also, change "Mean differences in PBL" to "Mean Difference in PBL" with the "s", and please add some text in the next paragraph so that I can understand what each value represents.}

\added{\textbf{Interactive Features.} The dashboard allows users to carry out different types of racial/ethnic and urbanization fairness analyses for each individual model. As shown in the four boxes at the bottom of Fig~\ref{fig:dashboard}, the dashboard includes the following key interactive features:}
\begin{itemize}
\setlength\itemsep{0.01em}
    \item \added{\textbf{Variables of Interest Selection}: Users can choose from six different analytical perspectives to analyze the fairness of a given COVID-19 forecast model: }
    \begin{itemize}
    \setlength\itemsep{0.01em}
        \item[--] \added{Race or ethnicity analysis, that allows to analyze the fairness of the predictions for a given minority race with respect to White (see example in Fig \ref{fig:dashboard})}
        \item[--] \added{Urbanicity analysis, that allow to evaluate the fairness of the predictions for a given urbanization level with respect to large metropolitan areas (see Fig 7 in the \nameref{s1appendix})}
        \item[--] \added{Analysis at the intersection of race/ethnicity and lookahead periods or pandemic phases, that allows to evaluate differences in COVID-19 forecast fairness for a given minority race (with respect to White) and for a given lookahead or pandemic phase (see Figs 8 \& 9 in \nameref{s1appendix} for a couple of examples)}
        \item[--] \added{Urbanicity and lookahead periods or pandemic phases analysis, allowing users to explore differences in COVID-19 forecast fairness for a given urbanization level (with respect to large metropolitan areas) and a given lookahead or phase (see Figs 10 \& 11 in \nameref{s1appendix} for a couple of examples)}

    \end{itemize}
    \item  \added{\textbf{Protected Variable Selection}: The dashboard allows users to focus on specific demographic groups:}
    \begin{itemize}
        \setlength\itemsep{0.01em}
        \item[--] \added{For racial/ethnic analysis: Black, Hispanic, or Asian AER (White is the baseline group)}
        \item[--] \added{For urbanicity analysis: Micropolitan or Small/Medium Metro AER (Large Metropolitan is the baseline group)}
    \end{itemize}
    \item \added{\textbf{Temporal Analysis Options}:}
    \begin{itemize}
        \setlength\itemsep{0.01em}
        \item[--] \added{Phase selection (0-6) for analyzing performance across different pandemic periods}
        \item[--] \added{Lookahead periods (7, 14, 21, or 28 days) for examining how prediction fairness varies with forecast horizon}
    \end{itemize}

\end{itemize}

%\vanessa{Saad, I know here we use Non-White because with the "majority" label we did not have enough Asian. However, I think that would be going into too many details here, I think? I suggest we just put "Asian" instead of Non-White in the text and in the plots. What do you think? NEW: I just changed to Asian in the text, and everything to plurality, you just need to change "non-white" aer to "asian aer" in figs 7,8 and 9}
%For example, Figure~\ref{fig:dashboard} shows an example of the dashboard for the exploration of individual model performance by race and type of model, with a focus on the relationship between the prediction errors for Hispanic and White counties ($AER_{Hispanic}$). The box plots for each model represent its AER distribution across all counties; and a user can explore the median AER as well as its quantiles for each predictive model. 
%Hovering over the model points displays all the information in the format of a `fairness nutritional card' as shown in Figure \ref{fig:fairness_card}, and inspired by the work of \cite{stoyanovich2019nutritional}.

\begin{figure}
\centering
\subfloat[\centering LUcompUncertLabVAR\_3streams\label{fig:LUcomp_model_fairness}]{\includegraphics[width=0.48\textwidth]{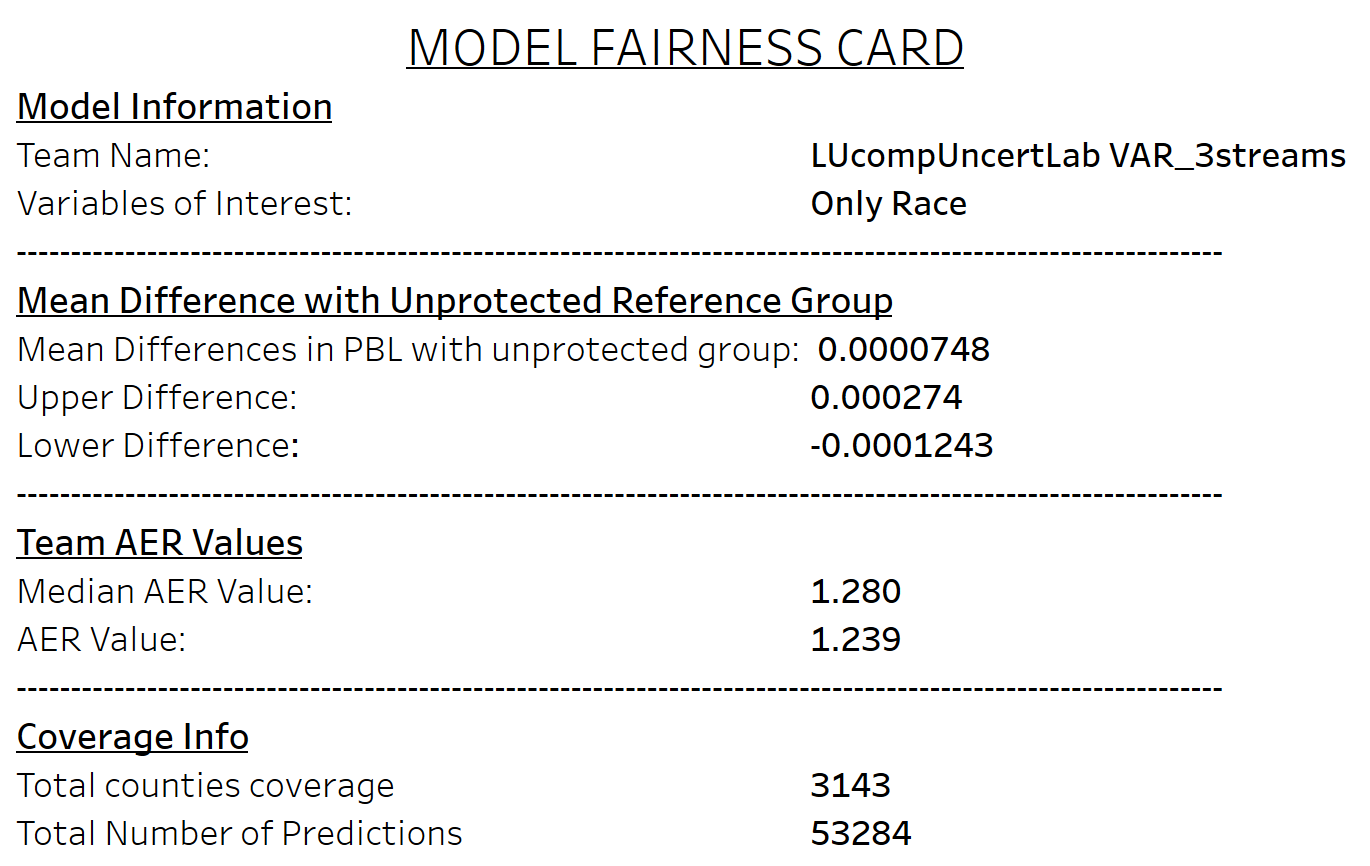}}
\centering
\subfloat[\centering IowaStateLW STEM\label{fig:iowa_model_fairness}]{\includegraphics[width=0.48\textwidth]{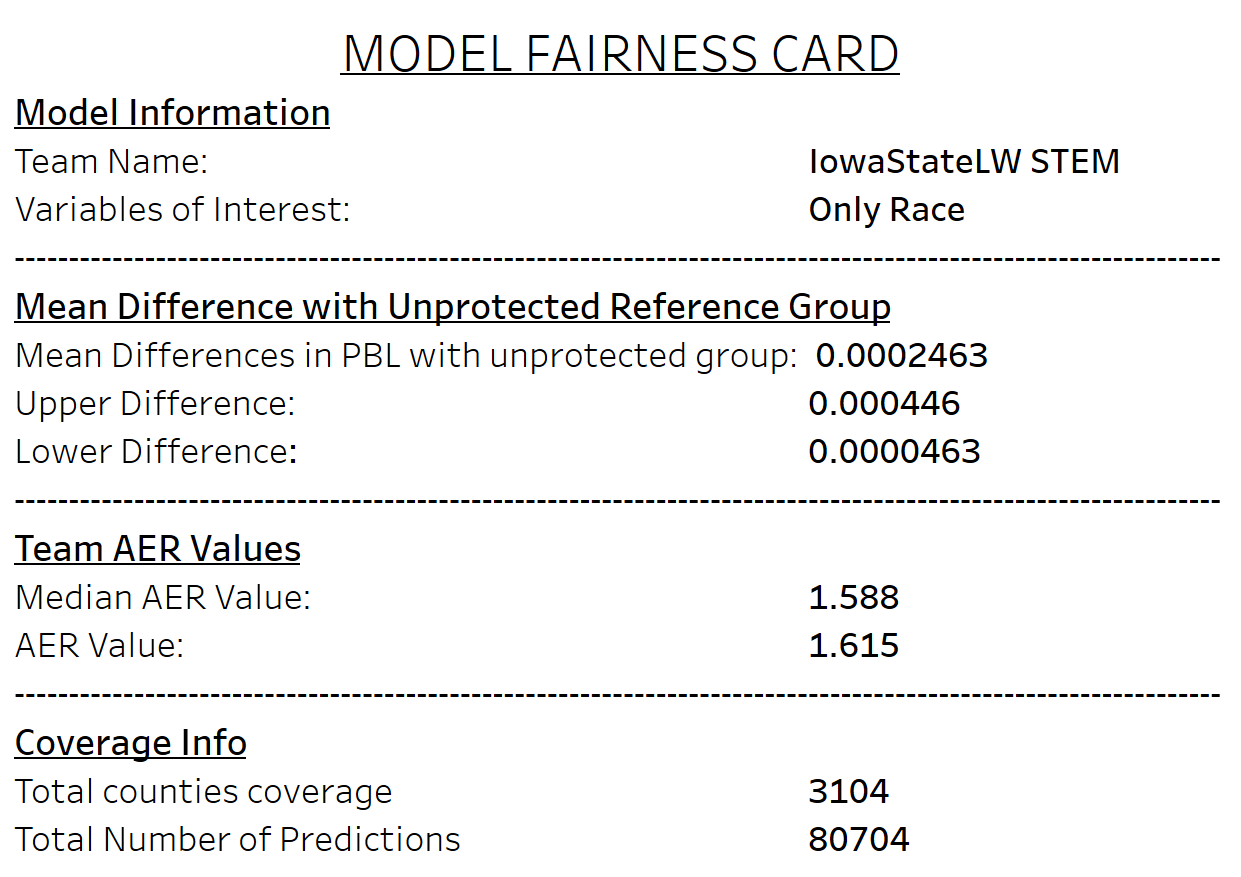}}
\caption{ Model fairness card displaying key performance metrics including model information, prediction error differences between protected and unprotected groups, AER values, and coverage statistics.}
\label{fig:fairness_card}
\end{figure}

\added{\textbf{Sample Use Case.} Fig~\ref{fig:dashboard} shows an example of the dashboard for the exploration of individual model performance by race and type of model, with a focus on the relationship between the prediction errors for Hispanic and White counties ($AER_{Hispanic}$). The box plots for each model represent its AER distribution across all counties; and a user can explore the median AER as well as its quantiles for each predictive model. In this example, most of the AERs are above one, pointing to unfair forecasts (higher errors) for the Hispanic group when compared to White counties. }

\added{Hovering over the model points displays all the information in the format of a `fairness nutritional card' as shown in Fig \ref{fig:fairness_card}. 'Nutritional labels' were proposed by Stoyanovich and Howe to assess model fairness \cite{stoyanovich2019nutritional} drawing an analogy to the food industry, where simple, standard labels convey information about the ingredients and production processes. We have adapted these cards to the COVID-19 fairness context, as a way to provide detailed COVID-19 forecast model fairness information. 
The fairness nutritional cards in our dashboard (see Fig \ref{fig:fairness_card} for a sample) provide detailed information organized into four key sections: (1) \textit{Model Information}, which identifies the team name and the variables being analyzed; (2) \textit{Mean Difference with Unprotected Reference Group}, which quantifies the prediction error differences between protected and unprotected groups in terms of PBL values, including upper and lower bounds; (3) \textit{Team AER Values}, showing both the median and specific AER values that indicate the relative performance between groups; and (4) Coverage Info, which provides context about the number of counties and total predictions covered by each team. }

%\textcolor{red}{Saad, I have added the paragraph below as an example, as we discussed. Can you please replace Fig 4 with two model cards, one for each of the models we discuss in the paragraph below? }

\added{Looking into Fig~\ref{fig:dashboard} and Fig \ref{fig:fairness_card}, we can observe that the \textit{IowaStateLW STEM} model exhibits a wide range of median AER values (min: 1.172, max: 4.772, median: 1.588) across its predictions, indicating highly variable fairness performance when assessed across different phases and lookaheads. On the other hand, the \textit{LUcompUncertLab VAR\_3streams} model shows consistent median AER values within a small range from 1.228 to 1.334 (median: 1.280). Comparing both models, we observe that although both of them are systematically producing predictions that are less fair for Hispanic counties (AER values are larger than 1), the  \textit{LUcompUncertLab VAR\_3streams} model has lower max AER values, suggesting that model might be a better choice. }

\added{Overall, we posit that the dashboard facilitates dynamic exploration of a diverse set of metrics across different dimensions, allowing users to examine how fairness measures vary with changes in pandemic phases and lookahead through the user interface. When exploring these temporal variables, the nutritional card automatically updates to display the relevant phase or lookahead information for the selected view.}

\section*{Discussion}

This study highlights the critical need to audit COVID-19 prediction models due to significant disparities in prediction accuracy. Our findings reveal that certain minority groups, especially Hispanic communities, and less urbanized areas consistently experience higher prediction errors. The race and ethnicity analysis revealed that increases in Hispanic population (when compared to White) exhibit significantly higher PBL errors; while increases in Asian and Black population are associated with lower PBL errors when compared to the White population (reference group), and while controlling for health outcomes and older population at the county level. 
The analysis on urbanization levels, on the other hand, revealed an inverse relationship between the level of urbanization and the magnitude of prediction errors underscoring the unique challenges encountered by rural areas. Rural counties consistently face higher prediction errors than their urban counterparts, a pattern that persists across various model types and forecast windows. 

%\vanessa{Saad, here are some thoughts on why for interactions, please see if you agree, edit}
\added{\textit{Potential Reasons.} These findings could be related to data quality (COVID-19 cases, mobility data) or structural problems. 
Hispanic groups or less urbanized areas being associated with significantly higher errors (when compared to their baselines: White population and large metropolitan area), could point to lack of quality COVID-19 case or mobility data for the Hispanic population and less urbanized areas. It could also be due to more complex spreading patterns that make COVID-19 cases harder to predict for the Hispanic population or rural areas, or two structural differences such as reduced access to medical facilities or testing sites. 
On the other hand, these findings are also pointing to better COVID-19 case or mobility data for Asian and Black population when compared to White; or to simpler spreading patterns that are easier to forecast, hence producing lower errors.} 

The implications of these findings are significant, since systematic disparities in model performance could lead to unfair distribution of public health resources or to less effective pandemic response efforts in Hispanic counties and in less densely populated regions, when compared to White and urban regions. 
Our definition of prediction fairness is focused on achieving similar prediction errors across racial, ethnic and urban-rural groups 
because COVID-19 cases have been used to make resource allocation and intervention decisions e.g., hospital beds or stay-at-home orders. Hence, higher prediction errors for minority racial groups or rural regions could in turn translate into unfair resource allocation for communities that have borne the brunt of the pandemic. Ultimately, we want to ensure our findings serve as a critical call to action for researchers and decision makers to analyze model performance disaggregated by racial/ethnic and urban-rural variables.

\added{
\textbf{Interaction Analysis}. 
Our interaction analysis provides a more multifaceted understanding of fairness in COVID-19 modeling. We find that deep learning models tend to produce the lowest disparities in errors across racial, ethnic and urban-rural groups, while 
compartmental and statistical models tend to be associated with the highest disparities. 
Our results have also shown that the use of mobility data helps reduce prediction error disparities for racial and ethnic groups as well as across urbanization levels. Short-term lookaheads and certain pandemic phases (case-peak phases 1, 3 and 4) are also associated with higher prediction error disparities for minority racial groups and rural areas. These findings highlight the complex interplay between model characteristics, data inputs, and social determinants in shaping prediction fairness. }

%\vanessa{Saad, here are some thought on why for interactions, please see if you agree, edit to form an argument}

\added{\textit{Potential Reasons.} Decreasing disparities in prediction errors for higher lookaheads could be due to a reduction in the effect of COVID-19 case data bias (positive or negative) for minority groups and less urbanized areas. 
In fact, as the lookahead increases, prior work has shown that case prediction becomes more difficult \cite{abrar2023analysis}, and this forecasting complexity appears to have a stronger effect than data bias on the prediction errors, thus
making all errors more similar (more fair) across racial/ethnic groups and urbanization levels. When looking into prediction error disparities across phases and race/ethnicity or urbanization levels, we argue that the higher prediction differences tend to take place during initial phases of the pandemic, which could point to factors like initial data collection issues, testing accessibility, or reporting practices that may have varied across racial and ethnic groups during the early stages of the pandemic, especially for Hispanic groups and less urbanized areas. It also revealing of the fact that there were very limited historical COVID-19 case data to learn from in the early phases.}

\added{Looking into types of models, deep learning (DL) models appear to be the best choice to reduce performance error differences between LMs and less urbanized areas; while DL models also appear to be a good compromise across minority groups. This could be pointing to DL modeling being able to better capture spatio-temporal dependencies without the behavioral assumptions of compartmental or statistical models, that show higher performance differences for some racial groups and across less urbanized levels.}

\added{Finally, mobility data appears to be providing additional information (insights into behavioral patterns) that helps reduce biases in model performance across the aforementioned sensitive groups. }

%(2) phases: we could argue that across minority race/ethnicity and less urbanized areas, we observe higher prediction differences during phases with higher volumes of cases, pointing to increases in (positive or negative) bias for minority races during phases where the pandemic spreading changes radically from past time. 

\added{
These interaction results highlight the need for  
%The findings of this study have significant implications for future COVID-19 modeling efforts and public health policy. M
researchers and modelers to carefully examine their data sources, model assumptions, and potential biases that could lead to unfair predictions for certain population groups. Incorporating fairness considerations into the model development, validation, and deployment processes is essential to ensure equitable outcomes. Public health officials and policymakers should be aware of the potential disparities in the accuracy of COVID-19 prediction models and work closely with modelers to mitigate these disparities. Failure to address these issues could lead to the perpetuation of health inequities and could eventually undermine the effectiveness of pandemic response efforts.}

\added{In addition to our findings, this study has several limitations that should be acknowledged. Firstly, we had to exclude some U.S. counties from our analysis due to insufficient data availability for the data sources we used. Second, while our work primarily focused on urbanicity and race/ethnicity as fairness-related variables, other important attributes, such as socioeconomic status or access to healthcare, etc, were not considered and could be explored in future research. Additionally, we were unable to incorporate all minority racial groups like AIAN and NHPI due to inadequate population sizes, which constrained our ability to assess fairness comprehensively across all demographic groups. }

\added{\textbf{Moving Forward.}} In our study, we have focused exclusively on county-level predictions because these are closer to local realities and allow for more actionable decision-making than state-level predictions. However, county-level statistics were collected only for COVID-19 cases, with hospitalizations or deaths only accounted for at the state level. Since prior work has shown that case counts might be more biased than hospitalization or death statistics \cite{Cramer2022-hub-dataset}, the results reflected in this paper could potentially change if hospitalization or death data were available at the county level and this study was replicated. 

%\saad{Saad: 1. add that county-level forecasts only report incident cases by the COVID-19 forecasthub, although literature suggests that the deaths and hospitalizations can be less biased. Additionally, county level forecasts give us a more granular view, that enables a more comprehensive view of prediction performances, particularly at a localized level. 
%2. We also have been constrained due to the lack of data at a county granularity
%for other relevant factors}

We posit that future research in COVID-19 case prediction models should focus on developing and validating bias mitigation strategies that account for performance disparities across race, ethnicity and urbanization levels. This may involve exploring alternative data sources, refining model architectures, and incorporating techniques to ensure fairness across different population groups. Additionally, more comprehensive and standardized race and ethnicity data collection in public health surveillance systems is crucial to enable accurate assessments of model fairness and to guide equitable decision-making.

\section*{Conclusions}

%Forecast Hub model predictions are shared by the CDC to support transparent decision making. Nevertheless, the performance of these models is measured only via accuracy, failing to report fairness metrics across critical social determinants during COVID-19 such as race, ethnicity and urbanization level. 
Our paper shows significant diverse predictive performance across social determinants for the Forest Hub COVID-19 models, with 
some minority racial and ethnic counties
%Asian, Black, Hispanic and Non-white counties 
as well as less urbanized counties often associated with statistically significant higher prediction errors. 
%When looking into model and data specifics, our results show that (i) compartmental models produce higher errors across protected groups, (ii) that using mobility data deepens unfairness across races, but not urbanization levels, (iii) that longer lookaheads are associated with higher PBL errors for protected groups and (iv) that early and recalibration pandemic phases increase the unfairness of the predictions across most protected groups. 
We also show that these higher errors are often times present for specific model types, lookaheads and pandemic phases;
and that these findings generally hold across different race associations.
\textbf{We hope this paper will encourage Forecast Hub modelers, the CDC and COVID-19 modelers to report fairness metrics together with accuracy, and to reflect on the potential negative impacts of the models on specific social groups and contexts. }

\section*{Acknowledgements}
\added{I would like to thank Dr. Peter Rankel for the statistical inputs for the paper. }

\added{We gratefully acknowledge the contributions of all research teams that have shared their models through the COVID-19 Forecast Hub, including Columbia University Projections (CU-Nochange, CU-scenario high, CU-scenario mid, CU-scenario low, CU-select) \cite{pei2020initial};
CMU-TimeSeries \cite{mcdonald2021beyond};
COVIDHub Models (COVIDhub CDC-ensemble, COVIDhub-trained ensemble, COVIDhub-ensemble, COVIDhub-4 week ensemble, COVIDhub-baseline) \cite{ray2020ensemble, ray2023comparing};
IEM MED-CovidProject \cite{IEM_MED_CovidProject};
LNQ-ens1 \cite{Wolfinge_Lander_2020};
JHU IDD-CovidSP \cite{lemaitre2021scenario};
CUBoulder-COVIDLSTM \cite{lucas2023spatiotemporal};
Google Harvard-CPF \cite{arik2020interpretable};
FRBSF Wilson-Econometric \cite{wilson2021weather};
UMass-MechBayes \cite{gibson2023real};
LANL-GrowthRate \cite{castro2021coffee};
UCLA-SuEIR \cite{zou2020epidemic};
CEID-Walk \cite{e3bo_random_walks};
RobertWalraven-ESG \cite{reichlab_covid19_forecast_hub_RobertWalraven_ESG};
OneQuietNight-ML \cite{Jo_Cho_OneQuietNight_COVID19Forecast};
IUPUI-HkPrMobiDyR \cite{chiang2022hawkes};
UpstateSU-GRU \cite{zhang2021seq2seq};
UChicagoCHATTOPADHYAY-UnIT \cite{huang2021universal};
PandemicCentral (PandemicCentral-USCounty, PandemicCentral-COVIDForest) \cite{galasso2022random};
JHUAPL-Bucky \cite{kinsey2020JHUAPL-Bucky};
LUcompUncertLab-VAR 3streams \cite{mcandrew_et_al_ComputationalUncertaintyLab};
FAIR-NRAR \cite{le2020neural};
JHU UNC GAS-StatMechPool \cite{HopkinsIDD_EpiForecastStatMech};
IowaStateLW-STEM \cite{wang2020spatiotemporal};
MITCovAlliance-SIR \cite{baek2021limits};
UVA-Ensemble \cite{adiga2023phase};}

\section*{Supporting Information}
\paragraph{S1 Appendix} \label{s1appendix} \textbf{This appendix provides supplementary information to
facilitate a deeper understanding and interpretation of the results.}

%\section*{Acknowledgments}
%Cras egestas velit mauris, eu mollis turpis pellentesque sit amet. Interdum et malesuada fames ac ante ipsum primis in faucibus. Nam id pretium nisi. Sed ac quam id nisi malesuada congue. Sed interdum aliquet augue, at pellentesque quam rhoncus vitae.

\nolinenumbers

% Either type in your references using
% \begin{thebibliography}{}
% Text
% \end{thebibliography}
%
% or
%
% Compile your BiBTeX database using our plos2015.bst
% style file and paste the contents of your .bbl file
% here. See http://journals.plos.org/plosone/s/latex for 
%\begin{thebibliography}{10}

\bibliography{ref}

\begin{thebibliography}{10}

\bibitem{forecasthub}
CDC. Forecast Hub; 2020.
\newblock \url{https://covid19forecasthub.org/}.

\bibitem{cdc}
CDC. CDC COVID-19 Visualization; 2020.
\newblock \url{https://www.cdc.gov/coronavirus/2019-ncov/science/forecasting/forecasting-math-modeling.html}.

\bibitem{Cramer2022-hub-dataset}
Cramer EY, Huang Y, Wang Y, Ray EL, Cornell M, Bracher J, et~al.
\newblock The United States COVID-19 Forecast Hub dataset.
\newblock Scientific Data. 2022;doi:{10.1101/2021.11.04.21265886}.

\bibitem{zhang2021seq2seq}
Zhang-James Y, Hess J, Salkin A, Wang D, Chen S, Winkelstein P, et~al.
\newblock A seq2seq model to forecast the COVID-19 cases, deaths and reproductive R numbers in US counties.
\newblock Research Square. 2021;.

\bibitem{arik2020interpretable}
Arik S, Li CL, Yoon J, Sinha R, Epshteyn A, Le L, et~al.
\newblock Interpretable sequence learning for COVID-19 forecasting.
\newblock Advances in Neural Information Processing Systems. 2020;33:18807--18818.

\bibitem{lucas2023spatiotemporal}
Lucas B, Vahedi B, Karimzadeh M.
\newblock A spatiotemporal machine learning approach to forecasting COVID-19 incidence at the county level in the USA.
\newblock International Journal of Data Science and Analytics. 2023;15(3):247--266.

\bibitem{le2020neural}
Le M, Ibrahim M, Sagun L, Lacroix T, Nickel M.
\newblock Neural relational autoregression for high-resolution COVID-19 forecasting.
\newblock Facebook AI Research. 2020;.

\bibitem{pei2020initial}
Pei S, Shaman J.
\newblock Initial simulation of SARS-CoV2 spread and intervention effects in the continental US.
\newblock MedRxiv. 2020; p. 2020--03.

\bibitem{zou2020epidemic}
Zou D, Wang L, Xu P, Chen J, Zhang W, Gu Q.
\newblock Epidemic model guided machine learning for COVID-19 forecasts in the United States.
\newblock MedRxiv. 2020; p. 2020--05.

\bibitem{chiang2022hawkes}
Chiang WH, Liu X, Mohler G.
\newblock Hawkes process modeling of COVID-19 with mobility leading indicators and spatial covariates.
\newblock International journal of forecasting. 2022;38(2):505--520.

\bibitem{galasso2022random}
Galasso J, Cao DM, Hochberg R.
\newblock A random forest model for forecasting regional COVID-19 cases utilizing reproduction number estimates and demographic data.
\newblock Chaos, Solitons \& Fractals. 2022;156:111779.

\bibitem{adiga2023phase}
Adiga A, Kaur G, Wang L, Hurt B, Porebski P, Venkatramanan S, et~al.
\newblock Phase-informed Bayesian ensemble models improve performance of COVID-19 forecasts.
\newblock In: Proceedings of the AAAI Conference on Artificial Intelligence. vol.~37; 2023. p. 15647--15653.

\bibitem{vieira2010querying}
Vieira MR, Frias-Martinez E, Bakalov P, Frias-Martinez V, Tsotras VJ.
\newblock Querying spatio-temporal patterns in mobile phone-call databases.
\newblock In: 2010 Eleventh International Conference on Mobile Data Management. IEEE; 2010. p. 239--248.

\bibitem{hernandez2017estimating}
Hernandez M, Hong L, Frias-Martinez V, Whitby A, Frias-Martinez E.
\newblock Estimating poverty using cell phone data: evidence from Guatemala.
\newblock World Bank Policy Research Working Paper. 2017;(7969).

\bibitem{frias2013cell}
Frias-Martinez V, Virseda J.
\newblock Cell phone analytics: Scaling human behavior studies into the millions.
\newblock Information Technologies \& International Development. 2013;9(2):pp--35.

\bibitem{rubio2010human}
Rubio A, Frias-Martinez V, Frias-Martinez E, Oliver N.
\newblock Human mobility in advanced and developing economies: A comparative analysis.
\newblock In: 2010 AAAI Spring Symposium Series; 2010.

\bibitem{wuspatial}
Wu J, Frias-Martinez E, Frias-Martinez V.
\newblock Spatial sensitivity analysis for urban hotspots using cell phone traces.
\newblock Environment and Planning B: Urban Analytics and City Science. 2021;.

\bibitem{frias2010socio}
Frias-Martinez V, Virseda J, Frias-Martinez E.
\newblock Socio-economic levels and human mobility.
\newblock In: Qual meets quant workshop-QMQ; 2010. p. 1--6.

\bibitem{fu2018identifying}
Fu C, McKenzie G, Frias-Martinez V, Stewart K.
\newblock Identifying spatiotemporal urban activities through linguistic signatures.
\newblock Computers, Environment and Urban Systems. 2018;72:25--37.

\bibitem{frias2012mobilizing}
Frias-Martinez V, Virseda J, Gomero A.
\newblock Mobilizing education: evaluation of a mobile learning tool in a low-income school.
\newblock In: Proceedings of the 14th international conference on Human-computer interaction with mobile devices and services; 2012. p. 441--450.

\bibitem{hong2016topic}
Hong L, Frias-Martinez E, Frias-Martinez V.
\newblock Topic models to infer socio-economic maps.
\newblock In: Proceedings of the AAAI Conference on Artificial Intelligence. vol.~30; 2016.

\bibitem{frias2012computing}
Frias-Martinez V, Soto V, Virseda J, Frias-Martinez E.
\newblock Computing cost-effective census maps from cell phone traces.
\newblock In: Workshop on pervasive urban applications; 2012.

\bibitem{wu2022enhancing}
Wu J, Abrar SM, Awasthi N, Frias-Martinez E, Frias-Martinez V.
\newblock Enhancing short-term crime prediction with human mobility flows and deep learning architectures.
\newblock EPJ Data Science. 2022;11(1):53.

\bibitem{wu2023auditing}
Wu J, Abrar SM, Awasthi N, Fr{\'\i}as-Mart{\'\i}nez V.
\newblock Auditing the fairness of place-based crime prediction models implemented with deep learning approaches.
\newblock Computers, Environment and Urban Systems. 2023;102:101967.

\bibitem{wesolowski2012quantifying}
Wesolowski A, Eagle N, Tatem AJ, Smith DL, Noor AM, Snow RW, et~al.
\newblock Quantifying the impact of human mobility on malaria.
\newblock Science. 2012;338(6104):267--270.

\bibitem{bengtsson2015using}
Bengtsson L, Gaudart J, Lu X, Moore S, Wetter E, Sallah K, et~al.
\newblock Using mobile phone data to predict the spatial spread of cholera.
\newblock Scientific reports. 2015;5(1):1--5.

\bibitem{hong2017understanding}
Hong L, Fu C, Torrens P, Frias-Martinez V.
\newblock Understanding citizens' and local governments' digital communications during natural disasters: the case of snowstorms.
\newblock In: Proceedings of the 2017 ACM on web science conference; 2017. p. 141--150.

\bibitem{isaacman2018modeling}
Isaacman S, Frias-Martinez V, Frias-Martinez E.
\newblock Modeling human migration patterns during drought conditions in La Guajira, Colombia.
\newblock In: Proceedings of the 1st ACM SIGCAS conference on computing and sustainable societies; 2018. p. 1--9.

\bibitem{ghurye2016framework}
Ghurye J, Krings G, Frias-Martinez V.
\newblock A framework to model human behavior at large scale during natural disasters.
\newblock In: 2016 17th IEEE International Conference on Mobile Data Management (MDM). vol.~1. IEEE; 2016. p. 18--27.

\bibitem{hong2020modeling}
Hong L, Frias-Martinez V.
\newblock Modeling and predicting evacuation flows during hurricane Irma.
\newblock EPJ Data Science. 2020;9(1):29.

\bibitem{erfani2023fairness}
Erfani A, Frias-Martinez V.
\newblock A fairness assessment of mobility-based COVID-19 case prediction models.
\newblock Plos one. 2023;18(10):e0292090.

\bibitem{badr2021limitations}
Badr HS, Gardner LM.
\newblock Limitations of using mobile phone data to model COVID-19 transmission in the USA.
\newblock The Lancet Infectious Diseases. 2021;21(5):e113.

\bibitem{abrar2023analysis}
Abrar SM, Awasthi N, Smolyak D, Frias-Martinez V.
\newblock Analysis of performance improvements and bias associated with the use of human mobility data in COVID-19 case prediction models.
\newblock ACM Journal on Computing and Sustainable Societies. 2023;.

\bibitem{gross2020racial}
Gross CP, Essien UR, Pasha S, Gross JR, Wang Sy, Nunez-Smith M.
\newblock Racial and ethnic disparities in population-level Covid-19 mortality.
\newblock Journal of general internal medicine. 2020;35:3097--3099.

\bibitem{souch2021commentary}
Souch JM, Cossman JS.
\newblock A commentary on rural-urban disparities in COVID-19 testing rates per 100,000 and risk factors.
\newblock The Journal of Rural Health. 2021;37(1):188.

\bibitem{tsai2022algorithmic}
Tsai TC, Arik S, Jacobson BH, Yoon J, Yoder N, Sava D, et~al.
\newblock Algorithmic fairness in pandemic forecasting: lessons from COVID-19.
\newblock NPJ Digital Medicine. 2022;5(1):59.

\bibitem{rajkomar2018ensuring}
Rajkomar A, Hardt M, Howell MD, Corrado G, Chin MH.
\newblock Ensuring fairness in machine learning to advance health equity.
\newblock Annals of internal medicine. 2018;169(12):866--872.

\bibitem{douglas2021variation}
Douglas MD, Respress E, Gaglioti AH, Li C, Blount MA, Hopkins J, et~al.
\newblock Variation in reporting of the race and ethnicity of COVID-19 cases and deaths across US states: April 12, 2020, and November 9, 2020.
\newblock American Journal of Public Health. 2021;111(6):1141--1148.

\bibitem{boehmer2002self}
Boehmer U, Kressin NR, Berlowitz DR, Christiansen CL, Kazis LE, Jones JA.
\newblock Self-reported vs administrative race/ethnicity data and study results.
\newblock American journal of public health. 2002;92(9):1471--1472.

\bibitem{cdcunderreporting}
CDC. Health Disparities; 2023.
\newblock Available from: \url{https://www.cdc.gov/nchs/nvss/vsrr/covid19/health_disparities.htm#CountyRaceHispanicOrigin}.

\bibitem{ama}
Del~Rios M, Puente S, Vergara-Rodriguez P, Sugrue N.
\newblock Invisibilidad de los latinos en la pandemia.
\newblock AMA Journal of Ethics. 2022; p. 289--295.

\bibitem{safegraph}
Safegraph. Safegraph Mobility Data; 2020.
\newblock \url{https://www.safegraph.com/guides/mobility-data}.

\bibitem{apple}
Apple. Apple Mobility Data; 2020.
\newblock \url{https://covid19.apple.com/mobility}.

\bibitem{google}
Google. Google Mobility Data; 2020.
\newblock \url{https://www.google.com/covid19/mobility/}.

\bibitem{ilin2021public}
Ilin C, Annan-Phan S, Tai XH, Mehra S, Hsiang S, Blumenstock JE.
\newblock Public mobility data enables COVID-19 forecasting and management at local and global scales.
\newblock Scientific reports. 2021;11(1):13531.

\bibitem{garcia2021improving}
Garc{\'\i}a-Cremades S, Morales-Garc{\'\i}a J, Hern{\'a}ndez-Sanjaime R, Mart{\'\i}nez-Espa{\~n}a R, Bueno-Crespo A, Hern{\'a}ndez-Orallo E, et~al.
\newblock Improving prediction of COVID-19 evolution by fusing epidemiological and mobility data.
\newblock Scientific Reports. 2021;11(1):15173.

\bibitem{coston2021leveraging}
Coston A, Guha N, Ouyang D, Lu L, Chouldechova A, Ho DE.
\newblock Leveraging administrative data for bias audits: Assessing disparate coverage with mobility data for COVID-19 policy.
\newblock In: Proceedings of the 2021 ACM Conference on Fairness, Accountability, and Transparency; 2021. p. 173--184.

\bibitem{gursoy2022error}
Gursoy F, Kakadiaris IA.
\newblock Error Parity Fairness: Testing for Group Fairness in Regression Tasks.
\newblock arXiv preprint arXiv:220808279. 2022;.

\bibitem{kader2021participatory}
Kader F, Smith CL.
\newblock Participatory approaches to addressing missing COVID-19 race and ethnicity data.
\newblock International Journal of Environmental Research and Public Health. 2021;18(12):6559.

\bibitem{stoyanovich2019nutritional}
Stoyanovich J, Howe B.
\newblock Nutritional labels for data and models.
\newblock A Quarterly bulletin of the Computer Society of the IEEE Technical Committee on Data Engineering. 2019;42(3).

\bibitem{acs}
CensusBureau. American Community Survey; 2023.
\newblock \url{https://www.census.gov/programs-surveys/acs}.

\bibitem{cdcurban}
CDC. CDC National Center for Health Statistics: Urban-Rural Classification Scheme for Counties; 2023.
\newblock \url{https://www.cdc.gov/nchs/data_access/urban_rural.htm\#2013_Urban-Rural_Classification_Scheme_for_Counties}.

\bibitem{centers2024underlying}
CDC, et~al.. Underlying conditions and the higher risk for severe COVID-19; 2024.

\bibitem{CDC_PLACES_2022}
CDC. {PLACES: Local Data for Better Health (Place Data 2022 Release)}; 2022.
\newblock \url{https://data.cdc.gov/500-Cities-Places/PLACES-Local-Data-for-Better-Health-Place-Data-202/eav7-hnsx/about_data}.

\bibitem{yang2020prevalence}
Yang J, Zheng Y, Gou X, Pu K, Chen Z, Guo Q, et~al.
\newblock Prevalence of comorbidities and its effects in patients infected with SARS-CoV-2: a systematic review and meta-analysis.
\newblock International journal of infectious diseases. 2020;94:91--95.

\bibitem{dong2020interactive}
Dong E, Du H, Gardner L.
\newblock An interactive web-based dashboard to track COVID-19 in real time.
\newblock The Lancet infectious diseases. 2020;20(5):533--534.

\bibitem{fox1992generalized}
Fox J, Monette G.
\newblock Generalized collinearity diagnostics.
\newblock Journal of the American Statistical Association. 1992;87(417):178--183.

\bibitem{fox2018r}
Fox J, Weisberg S.
\newblock An R companion to applied regression.
\newblock Sage publications; 2018.

\bibitem{castelnovo2022clarification}
Castelnovo A, Crupi R, Greco G, Regoli D, Penco IG, Cosentini AC.
\newblock A clarification of the nuances in the fairness metrics landscape.
\newblock Scientific Reports. 2022;12(1):4209.

\bibitem{mcdonald2021beyond}
McDonald DJ, Bien J, Green A, Hu AJ, DeFries N, Hyun S, et~al.
\newblock Beyond Cases and Deaths: The Benefits of Auxiliary Data Streams In Tracking the COVID-19 Pandemic: Can auxiliary indicators improve COVID-19 forecasting and hotspot prediction?
\newblock Proceedings of the National Academy of Sciences of the United States of America. 2021;118(51).

\bibitem{ray2020ensemble}
Ray EL, Wattanachit N, Niemi J, Kanji AH, House K, Cramer EY, et~al.
\newblock Ensemble forecasts of coronavirus disease 2019 (COVID-19) in the US.
\newblock MedRXiv. 2020; p. 2020--08.

\bibitem{ray2023comparing}
Ray EL, Brooks LC, Bien J, Biggerstaff M, Bosse NI, Bracher J, et~al.
\newblock Comparing trained and untrained probabilistic ensemble forecasts of COVID-19 cases and deaths in the United States.
\newblock International journal of forecasting. 2023;39(3):1366--1383.

\bibitem{IEM_MED_CovidProject}
Suchoski B, Stage S, Gurung H, Baccam S. IEM\_MED-CovidProject;.
\newblock \url{https://iem-modeling.com/}.

\bibitem{Wolfinge_Lander_2020}
Wolfinge R, Lander D. LockNQuay -- LNQ-ens1; 2020.
\newblock Kaggle, \url{https://www.kaggle.com/sasrdw/locknqua}.

\bibitem{lemaitre2021scenario}
Lemaitre JC, Grantz KH, Kaminsky J, Meredith HR, Truelove SA, Lauer SA, et~al.
\newblock A scenario modeling pipeline for COVID-19 emergency planning.
\newblock Scientific reports. 2021;11(1):7534.

\bibitem{wilson2021weather}
Wilson DJ. Weather, mobility, and COVID-19: a panel local projections estimator for understanding and forecasting infectious disease spread. Federal Reserve Bank of San Francisco. February 2021; 2021.

\bibitem{gibson2023real}
Gibson GC, Reich NG, Sheldon D.
\newblock Real-time mechanistic bayesian forecasts of covid-19 mortality.
\newblock The annals of applied statistics. 2023;17(3):1801.

\bibitem{castro2021coffee}
Castro L, Fairchild G, Michaud I, Osthus D.
\newblock COFFEE: COVID-19 forecasts using fast evaluations and estimation.
\newblock arXiv preprint arXiv:211001546. 2021;.

\bibitem{e3bo_random_walks}
e3bo. random-walks;.
\newblock \url{https://github.com/e3bo/random-walks}.

\bibitem{reichlab_covid19_forecast_hub_RobertWalraven_ESG}
reichlab. COVID-19 Forecast Hub -- RobertWalraven-ESG;.
\newblock \url{https://github.com/reichlab/covid19-forecast-hub/tree/master/data-processed/RobertWalraven-ESG}.

\bibitem{Jo_Cho_OneQuietNight_COVID19Forecast}
Jo A, Cho J. One-Quiet-Night COVID-19-forecast;.
\newblock \url{https://github.com/One-Quiet-Night/COVID-19-forecast}.

\bibitem{huang2021universal}
Huang Y, Chattopadhyay I.
\newblock Universal risk phenotype of US counties for flu-like transmission to improve county-specific COVID-19 incidence forecasts.
\newblock PLoS computational biology. 2021;17(10):e1009363.

\bibitem{kinsey2020JHUAPL-Bucky}
Kinsey M, Tallaksen K, Obrecht RF, Asher L, Costello C, Kelbaugh M, et~al.. JHUAPL-Bucky; 2020.
\newblock \url{https://github.com/mattkinsey/bucky}.

\bibitem{mcandrew_et_al_ComputationalUncertaintyLab}
McAndrew T, Piriya M, Berlin A, Gandhi PD. Computational Uncertainty Lab;.
\newblock \url{https://github.com/computationalUncertaintyLab}.

\bibitem{HopkinsIDD_EpiForecastStatMech}
HopkinsIDD. Exploring methods for merging mechanistic and statistical models to forecast epidemics;.
\newblock \url{https://github.com/HopkinsIDD/EpiForecastStatMech}.

\bibitem{wang2020spatiotemporal}
Wang L, Wang G, Gao L, Li X, Yu S, Kim M, et~al.
\newblock Spatiotemporal dynamics, nowcasting and forecasting of COVID-19 in the United States.
\newblock arXiv preprint arXiv:200414103. 2020;.

\bibitem{baek2021limits}
Baek J, Farias VF, Georgescu A, Levi R, Peng T, Sinha D, et~al.
\newblock The limits to learning a diffusion model.
\newblock In: Proceedings of the 22nd ACM Conference on Economics and Computation; 2021. p. 130--131.

\end{thebibliography}
%\bibitem{bib1}
%Conant GC, Wolfe KH.
%\newblock {{T}urning a hobby into a job: how duplicated genes find new
%  functions}.
%\newblock Nat Rev Genet. 2008 Dec;9(12):938--950.

%\bibitem{bib2}
%Ohno S.
%\newblock Evolution by gene duplication.
%\newblock London: George Alien \& Unwin Ltd. Berlin, Heidelberg and New York:
%  Springer-Verlag.; 1970.

%\bibitem{bib3}
%Magwire MM, Bayer F, Webster CL, Cao C, Jiggins FM.
%\newblock {{S}uccessive increases in the resistance of {D}rosophila to viral
%  infection through a transposon insertion followed by a {D}uplication}.
%\newblock PLoS Genet. 2011 Oct;7(10):e1002337.

%\end{thebibliography}

% Include only the SI item label in the paragraph heading. Use the \nameref{label} command to cite SI items in the text.

\end{document}